\documentclass[aps,pre,twocolumn,groupedaddress]{revtex4-1}

\usepackage{lipsum}
\usepackage{graphicx}
\usepackage{amsmath,amssymb}
\usepackage{bm}
\usepackage{dcolumn}
\usepackage{theorem}
\usepackage{wrapfig}
 \usepackage{mathtools}
  \usepackage{xcolor}
 \usepackage{txfonts}

\mathtoolsset{centercolon}

\theoremstyle{break}

\newcounter{mycounter}
\newcommand{\Remark}[1][]{\refstepcounter{mycounter}#1\textit{Remark} \arabic{mycounter}. }

\newcommand{\article}[7]{{#1}: {#3} {\bf #4}, {#5} ({#7}).}
\newcommand{\book}[4]{{#1}: {\it #2} ({#3}, {#4})} 
 
\newcommand{\wip}[3]{{\it #1}, ({#2}, {#3})} 

\newcommand{\ftrunc}[1]{f_{({#1})}}
\newcommand{\ftruncE}[1]{\ftrunc{N}^\equilibrium}
\newcommand{\cc}{\mathbf{c}}
\newcommand{\xx}{\mathbf{x}}

\newcommand{\R}{\varmathbb{R}}
\newcommand{\inta}{\int_{\R^3} \int_0^\infty \int_0^\infty}
\newcommand{\da}{\, dI^R dI^V d\cc}

 \newcommand{\pr}{\varphi\left(I^R\right)}
\newcommand{\pv}{\psi\left(I^V\right)}

\newcommand{\state}{Statement}
\newtheorem{theorem}{\state}

\makeatletter

\begin{document}


\title{Rational Extended Thermodynamics of a Rarefied Polyatomic Gas \\ with Molecular Relaxation Processes} 


\author{Takashi Arima$^1$, Tommaso Ruggeri$^2$, and Masaru Sugiyama$^3$}
\email[]{arima@kanagawa-u.ac.jp, tommaso.ruggeri@unibo.it, sugiyama@nitech.ac.jp}
\affiliation{$^1$Department of Mechanical Engineering, Faculty of Engineering, Kanagawa University, Yokohama 221-8686,  Japan\\
$^2$Department of Mathematics, University of Bologna, Bologna, Italy\\
$^3$Nagoya Institute of Technology, Nagoya 466-8555, Japan}
\date{\today}

\begin{abstract}
We present a more refined version of rational
extended thermodynamics   of rarefied polyatomic gases in which molecular rotational and vibrational relaxation processes are treated individually.  In this case we need a triple hierarchy of the moment system and the system of balance equations is closed via the maximum entropy principle.  Three different types of the production terms in the system, which are suggested by a generalized BGK-type collision term in the Boltzmann equation, are adopted.  In particular, the rational extended thermodynamic theory with seven independent fields (ET$_7$) is analyzed in detail.  Finally, the dispersion relation of ultrasonic wave derived from the ET$_7$ theory is confirmed by the experimental data for CO$_2$, Cl$_2$, and Br$_2$ gases.
\end{abstract}



\maketitle

\section{Introduction}

Nonequilibrium phenomena observed in polyatomic gases, where energy exchanges among the translational, rotational, and vibrational modes of a molecule play a key role \cite{HerzfeldRice}, have attracted longstanding interest in various fields such as physics, chemistry, engineering. To describe such phenomena, the thermodynamic theory of relaxation processes of internal variables \cite{Mandelstam-1937,Meixner-1943,Meixner-1952}, which can be set within the framework of thermodynamics of irreversible processes (TIP) \cite{de Groot}, has been adopted.  Absorption and dispersion of ultrasonic waves \cite{Herzfeld,Bhatia}, and shock waves \cite{Vincenti}, in particular, have been studied by using the theory.

TIP relies essentially on the assumption of local equilibrium \cite{de Groot}. 
A theory of viscous heat-conducting fluids based on TIP is the well-known Navier-Stokes Fourier theory of the Newtonian fluids. 
Nowadays, however, there exist increasing demands for deeper understanding of strong nonequilibrium phenomena in polyatomic gases, that is, the phenomena out of local equilibrium in nano-technology, space science, molecular biology, and so on \cite{ET,RET,book,JouBook,Lebon,Grmela2011,Grmela2016}.   

Rational extended thermodynamics (hereafter referred to as ET for simplicity instead of RET) \cite{ET,RET,book} has been developed as a thermodynamic theory being applicable to nonequilibrium phenomena with steep gradients and rapid changes in space-time, which are out of local equilibrium.  ET of rarefied monatomic gases is summarized in \cite{ET,RET}, while ET of rarefied polyatomic gases with one relaxation process is presented in \cite{book}.  In ET, two different closure methods of a system of field equations have been proposed and extensively applied to various problems:
\begin{itemize}
\item  \emph{Phenomenological ET}: The closure is obtained by using the universal principles of continuum thermomechanics -- objectivity, entropy, and causality principles -- to select admissible constitutive equations (see \cite{LiuMul}, \cite{ET}, \cite{RET} for monatomic gases and \cite{Arima-2011}, \cite{book} for polyatomic ones);
\item   \emph{Molecular ET}: The fields are  moments of a distribution function and the closure is obtained by using the maximum entropy principle (MEP) \cite{Dreyer-1987,ET}. 
In molecular ET, it was proved that the closure by the MEP is equivalent to imposition of the entropy principle on the truncated moment equations both for monatomic gases \cite{Boillat-1997}
and for polyatomic gases \cite{Arima-2014}.
\end{itemize}
It was verified that the two closure methods are equivalent to each other and also equivalent to the Grad kinetic closure based on the perturbation around the Maxwellian via  Hermite polynomials \cite{Grad} (see \cite{Dreyer-1987,ET,RET} for monatomic gases with $13$ fields and   \cite{Pavic-2013,book,Mallinger} for polyatomic gases with $14$ fields).
  
  \bigskip
  
For later reference, we briefly explain ET of rarefied polyatomic gases with one relaxation process \cite{book}.
In polyatomic gases, the molecular internal degrees of freedom, which are not present in monatomic gases, come into play \cite{Kremer-2010}.  In particular, the internal specific energy is no longer related to the pressure in a simple way.      

A phenomenological ET theory with the binary hierarchy was firstly established by Arima, Taniguchi, Ruggeri and Sugiyama \cite{Arima-2011}, where $14$ independent fields: mass density, velocity, specific internal energy, shear stress, dynamic (nonequilibrium) pressure, and heat flux are adopted.  This theory is called ET$_{14}$.  The Navier-Stokes Fourier theory is included in ET$_{14}$ as a limiting case.  
  
Concerning its kinetic counterpart, a crucial step towards the development 
of the theory of rarefied polyatomic gases was made by an idea of Borgnakke and Larsen \cite{Borgnakke-1975}.
The distribution function is assumed to depend on an additional continuous variable 
representing the energy of the internal degrees of a molecule in order to take into account the exchange 
of energy (other than translational one) in binary collisions. This model was initially used 
for Monte Carlo simulations of polyatomic gases, and later it was applied to the derivation 
of the generalized Boltzmann equation by Bourgat, Desvillettes, Le Tallec, and Perthame
\cite{Bourgat-1994}, and 
was applied also to chemically reacting mixtures \cite{Desvillettes-2005}.
  
In this model, a non-negative energy of the internal degrees of a molecule, $I$, is introduced. The velocity distribution function depends on this additional parameter, i.e., $f \equiv f\left(\xx, \cc, t, I\right)$, where $f(\xx,\cc, t, I)\, d\xx\, d\cc$ is the number density of molecules with the energy $I$ at time $t$ and in the volume element $d\xx\, d\cc$ of the phase space (6D position-velocity space) centered at $\left(\xx,\cc\right) \in \R^3\times \R^3$. 
The Boltzmann equation is formally the same as the one of monatomic gases:
\begin{equation} 
\partial_t f  + c_i\, \partial_i f = Q\left(f\right), \label{eq:Boltzmann}
\end{equation}
but, for the collision term $Q(f)$, we take into account the influence of internal degrees of freedom through the collision cross-section \cite{Borgnakke-1975,Bourgat-1994}.  Here $\partial_t \equiv \partial / \partial t$ and $\partial_i \equiv \partial / \partial x_i$.
Then, from the Boltzmann equation \eqref{eq:Boltzmann}, we have  a binary hierarchy  of the field equations \cite{Pavic-2013,Arima-2014,book}: 
\begin{align} \label{eq:hierarchy-poly}
& \partial_t F + \partial_i F_i = 0,  \notag \\
& \partial_t F_{i_1} + \partial_i F_{ii_1} = 0, \notag \\
& \partial_t F_{i_1 i_2} + \partial_i F_{ii_1 i_2} = {P_{i_1 i_2}},
&& \partial_t G_{ll} + \partial_i G_{lli} = {0},\\
& \partial_t F_{i_1 i_2 i_3} + \partial_i F_{ii_1 i_2 i_3} = {P_{i_1 i_2 i_3}},
&& \partial_t G_{lli_1} + \partial_i G_{lli i_1} = {Q_{lli_1}}, \notag\\
& \qquad \ \vdots  && \qquad \ \vdots \notag 
\end{align}
involving the \emph{momentum-like} moments $F$ and the \emph{energy-like} moments $G$:  
\begin{align*}
   &F = \int_{\R^3} \int_0^\infty m f  \phi(I)\, dI d\cc,\\
  &F_{i_{1}\ldots i_{j}} = \int_{\R^3} \int_0^\infty m  c_{i_1}\cdots c_{i_j} f \phi(I)\, dI d\cc, \\
  &G_{ll} = \int_{\R^3} \int_0^\infty m  \left(c^2+\frac{2I}{m}\right) f \phi(I)\, dI d\cc,\\
  &G_{lli_{1}\ldots i_{k}} = \int_{\R^3} \int_0^\infty m  \left(c^2+\frac{2I}{m}\right) c_{i_1}\cdots c_{i_k} f \phi(I)\, dI d\cc, 
\end{align*}
where $m$ is the mass of a molecule, $\phi(I)$ is the state density of the internal mode, that is, $\phi(I)dI$ represents the number of the internal states of a molecule having the internal energy between $I$ and $I+dI$, and $j, k=1,2, \cdots$.  The first five moments are conserved quantities: the mass density $F (=\rho)$, the momentum density $F_i (=\rho v_i)$, and twice the energy density $G_{ll} (=2\rho \varepsilon + \rho v^2 = 2\rho (\varepsilon^K + \varepsilon^I) + \rho v^2)$, where $v_i$ is the mean velocity (and $v^2 = v_i v_i$), and $\varepsilon$ is the specific internal energy composed of the kinetic part $\varepsilon^K$ and the internal part $\varepsilon^I$.  The quantities $P$'s and $Q$'s in the right hand side of \eqref{eq:hierarchy-poly} are the production terms derived from the collision term:
\begin{align*}
& P_{i_{1}\ldots i_{j}} = \int_{\R^3} \int_0^\infty m  c_{i_1}\cdots c_{i_j} Q(f) \phi(I)\, dI d\cc, \\
& Q_{lli_{1}\ldots i_{k}} = \int_{\R^3} \int_0^\infty m  \left(c^2+\frac{2I}{m}\right) c_{i_1}\cdots c_{i_k} Q(f) \phi(I)\, dI d\cc, 
\end{align*}
where $j=2,3, \cdots$ and $k=1,2, \cdots $.

Using the molecular approach and the MEP, Pavi\'c, Ruggeri, and Simi\'c  \cite{Pavic-2013} (see also \cite{book})
deduced the equilibrium distribution function that maximizes the entropy:
\begin{align}
 \bar{f}_E = \frac{\rho}{m A(T)}\left(\frac{m}{2\pi k_B T}\right)^{3/2} \exp \left\{- \frac{1}{k_B T}\left(\frac{1}{2}mC^2 + I\right)\right\}, \label{fEone}
\end{align}
which is the generalized Maxwellian in the case of polyatomic gases.
$A(T)$ is the normalization factor:
\begin{align*}
  A(T) = \int_0^\infty \phi(I) \mathrm{e}^{- {\beta_E I }}\mathrm{d}I,   
   \label{AEione}
\end{align*}
where $\beta_E \equiv 1/(k_B T)$, $k_B$ is the Boltzmann constant, $T$ is the absolute temperature related with the kinetic energy in equilibrium:
\begin{equation*}
\varepsilon^K_E = \frac{3}{2}\frac{k_B}{m}T,
\end{equation*}
and $C^2 = C_i C_i$ with $C_i \equiv c_i - v_i$ being the peculiar velocity. 
Then the same authors derived the system of ET$_{14}$ using the MEP and 
obtained the same closure as the one in the phenomenological approach \cite{Arima-2011}.

The validity of ET$_{14}$ has been confirmed by comparing the theoretical predictions to the experimental data of linear waves \cite{ET14linear}, shock waves \cite{ET14shock,Kosuge}, and light scattering \cite{Arima-2013f}, in particular, in the region where the Navier-Stokes Fourier theory fails.

If all the dissipative fluxes except for the dynamic pressure are negligible, ET$_{14}$ reduces to a simpler ET theory with six independent fields (ET$_6$): mass density, velocity, specific internal energy, and dynamic pressure \cite{ET6,ET6Meccanica}.  This theory is the simplest extension of the Euler theory of perfect fluids and is compatible with the Meixner theory with one internal variable \cite{Meixner-1943,Meixner-1952}.  The correspondence relation between ET$_6$ and the Meixner theory was shown explicitly in \cite{ET6}.  The distinct shock wave structure observed in polyatomic gases such as CO$_2$ gas is explained satisfactorily also by the ET$_6$ theory \cite{ET6shock}.  

Furthermore the ET$_6$ theory with a nonlinear constitutive equation was studied in detail \cite{ET6nonlinear,ET6nonlinearshock,Wascom2015a,Wascom2015b}. It is noteworthy that the nonlinear  ET$_6$ theory is perfectly consistent with the molecular approach of the kinetic theory in polytropic gases \cite{Ruggeri-2016} and also in non-polytropic ones \cite{RuggeriSpiga}. In particular, in \cite{RuggeriSpiga}, comparison was also made between the present method via the continuous energy parameter $I$ in the distribution function and the mixture-like approach based on a discrete internal energy given by Groppi and Spiga \cite{GS}.

The ET theory with any number of independent fields has also been constructed \cite{Arima-2014,Arima-2014b}, and the convergence to the singular limit of monatomic gas when the degrees of freedom of a molecule $D\to 3$ was proved \cite{Arima-2013,Arima-2016}.

It is evident, however, that the ET theory of polyatomic gases with the binary hierarchy has the limitation of its applicability, although the theory has been successfully utilized to analyze various nonequilibrium phenomena as explained above.  In fact, we have many experimental data showing that the relaxation times of the rotational mode and of the vibrational mode are quite different to each other.  In such a case, more than one molecular relaxation processes should be taken into account to make the ET theory more precise.  Our aim of the present paper is to establish such an ET theory with much wider applicability range for rarefied polyatomic gases and to show its usefulness by studying ultrasonic wave propagation.

The present paper is organized as follows: In Section \ref{sec:2DistFun}, we explain the kinetic model for a polyatomic gas with two internal relaxation processes by using two parameters expressing the rotational and vibrational energies of a molecule. The equilibrium distribution function, the expressions of the thermal and caloric equations of state, and the entropy density in equilibrium are also shown.
In Section \ref{sec:3Noneq}, we make a general discussion on the system of balance equations in ET of polyatomic gases.  Defining three kinds of moments, we derive a triple hierarchy of moment equations from the Boltzmann equation.  And we study the truncated system of balance equations and its closure via MEP.
In Section \ref{sec:prodmol}, we introduce a simple collision term with three relaxation times, which is a generalization of the BGK-model.
In Section \ref{sec:ET7}, we establish the ET$_7$ theory with seven independent fields: mass density, momentum density, translational energy density, rotational energy density, and vibrational energy density.   
We derive the nonequilibrium distribution function and the closed system of field equations. 
In Section \ref{sec:FeaturesET7}, we summarize some features of the ET$_7$ theory.
In Section \ref{sec:LW}, we study the dispersion relation of a plane harmonic wave.  Theoretical prediction of the attenuation is compared with the experimental data for CO$_2$, Cl$_2$ and Br$_2$ gases. 
Final section is devoted to the concluding remarks and the discussion on some future problems.

\section{Distribution function with two energies of internal modes}\label{sec:2DistFun}

We adopt the closure of molecular ET in this paper, therefore we firstly explain the kinetic model for a polyatomic gas with two internal relaxation processes and then derive its equilibrium distribution function.  The thermal and caloric equations of state, and the expression of the entropy density in equilibrium are also shown.

In order to describe the relaxation processes of rotational and vibrational modes separately, we decompose the energy of internal modes $I$ as the sum  of  the energy of rotational mode $I^R$ and the energy of vibrational mode $I^V$:  
\begin{align}
I= I^R + I^V.
\label{decompose}
\end{align}
Generalizing the Borgnakke-Larsen idea \cite{Borgnakke-1975}, we assume the same form of the Boltzmann equation \eqref{eq:Boltzmann} with a velocity distribution function that depends on these additional parameters, i.e., $f \equiv f \left(\xx, \cc, t, I^R, I^V\right)$. And we also take into account the effect of the parameters $I^R$ and $I^V$ on the collision term $Q(f)$.

  \medskip
  
\Remark
As a state near the dissociation temperature, in which the molecular vibration is highly anharmonic, is out of the scope of the present study, the relation \eqref{decompose} can be safely assumed.

\Remark[\label{R:modes}] In a harmonic approximation of the molecular vibration, we may further divide $I^V$ into the energies of several harmonic modes.  
However, in this paper, as we focus our study on the contribution from the rotational or vibrational mode as a whole, we do not enter into such details although the generalization in this direction is straightforward.

\subsection{Equilibrium distribution function}
 
We derive the equilibrium distribution function $f_E$ by means of MEP. 
We remark that the collision invariants of the present model are $m$, $m c_i$, and $m c^2+2I^R+2I^V$.  
These quantities correspond to the hydrodynamics variables, i.e., the mass density $F(=\rho)$,  the momentum density $F_i(=\rho v_i)$ and twice the energy density $G_{ll} (= 2\rho \varepsilon + \rho v^2)$ through the following relations:
\begin{align}
	\begin{split}
   &F =  \inta m f \, \pr \pv \da, \qquad\\
   &F_i =  \inta m c_i f \, \pr \pv \da,\\
	 &G_{ll}  = \inta \left(m c^2 + 2I^R + 2I^V\right)f \, \pr \pv \da.
	\end{split}
  \label{hydro}
\end{align}
Here $\pr$ and $\pv$ are the state densities corresponding to $I^R$ and $I^V$.  And it is easy to see from \eqref{hydro}$_3$, that the specific internal energy $\varepsilon$ is composed of the kinetic part $\varepsilon^K$ and the parts of rotational mode $\varepsilon^R$ and of vibrational mode $\varepsilon^V$, i.e., 
\begin{equation*}
\varepsilon = \varepsilon^K + \varepsilon^R + \varepsilon^V.
\end{equation*}

The entropy density $h$ is defined by
\begin{equation} \label{entropy}
	h = - k_B \inta f \log f\, \pr \pv \da,
\end{equation}
where $k_B$ is the Boltzmann constant.

\begin{theorem}\label{feq}
  The equilibrium distribution function $f_E$, which maximizes the entropy density \eqref{entropy} under the constraints \eqref{hydro}, is given by
\begin{align}
 f_E = \frac{\rho}{m A^R(T)A^V(T)}\left(\frac{m}{2\pi k_B T}\right)^{3/2} \exp \left\{- \frac{1}{k_B T}\left(\frac{1}{2}mC^2 + I^R + I^V\right)\right\}, \label{fE}
\end{align}
where $A^R(T)$ and $A^V(T)$ are normalization factors:
\begin{align}
  A^R(T) = \int_0^\infty \pr \mathrm{e}^{- {\beta_E I^R}}\mathrm{d}I^R, \ \ 
  A^V(T) = \int_0^\infty \pv \mathrm{e}^{- {\beta_E I^V}}\mathrm{d}I^V. 
   \label{AEi}
\end{align}
\end{theorem}

The proof is omitted here, for simplicity, because it is essentially the same as the one shown in \cite{Pavic-2013,Ruggeri-2016,book}. In fact, replacing $I$, $A(T)$ and $\phi\left(I\right)$ in \eqref{fEone} by $I^R+I^V$, $\pr \pv $, and $A^R(T) A^V(T)$, respectively, we can obtain \eqref{fE}.

The equilibrium distribution function can be expressed by the product of the equilibrium distribution functions of the three modes:
\begin{align*}
f_E=f^{(K)}_E f^{(R)}_E f^{(V)}_E, 
\end{align*}
where
\begin{align*}
&f^{(K)}_E = \frac{\rho}{m}\left(\frac{m}{2\pi k_B T}\right)^{3/2} \exp \left(- \frac{mC^2}{2 k_B T}\right), \\
&f^{(R)}_E = \frac{1}{A^R(T)} \exp \left(- \frac{I^R}{k_B T}\right), \quad
f^{(V)}_E = \frac{1}{A^V(T)} \exp \left(- \frac{I^V}{k_B T}\right).
\end{align*}

\subsection{Thermal and caloric equations of state}

By using the equilibrium distribution function $f_E$, we obtain the thermal and caloric equations of state.  The pressure $p$ is expressed by
\begin{align}
p = p^{K}(\rho, T) \equiv \frac{k_B}{m}\rho T. \label{ThEqSt}
\end{align}
The caloric equation of state is given by
\begin{align}
\varepsilon = \varepsilon_E(T) = \varepsilon^{K}_E(T) + \varepsilon^R_E(T) + \varepsilon^V_E(T),  \label{CaEqSt}
\end{align}
and, proceeding in similar way as shown in \cite{RuggeriSpiga}, we have
\begin{align}
 \begin{split}
	&\varepsilon^K_E(T) \equiv \frac{3}{2}\frac{k_B}{m}T, \\
    &\varepsilon^R_E(T) \equiv \frac{k_B}{m}T^2\frac{\mathrm{d} \log A^R(T)}{\mathrm{d}T} ,\\
    &\varepsilon^V_E(T) \equiv \frac{k_B}{m}T^2\frac{\mathrm{d} \log A^V(T)}{\mathrm{d}T} .    
 \end{split}
  \label{int1e}
\end{align}
Therefore if we know the normalization factors $A^R(T)$ and $A^V(T)$, similar to the partition function in statistical mechanics, we can derive the equilibrium energies of rotational and vibrational modes from \eqref{int1e}. Vice versa if we know, at the  macroscopic phenomenological level, the constitutive equations   $\varepsilon^R_E(T)$ and $\varepsilon^V_E(T)$, we can obtain by  integration of \eqref{int1e}$_{2,3}$
\begin{align*}
 &A^R(T) = A^R_0  \exp\left(\frac{m}{k_B}\int_{T_0}^{T} \frac{\varepsilon^R_T({T^\prime}  )}{{T^\prime}  ^2} d{T^\prime}   \right), \\
 &A^V(T) = A^V_0  \exp\left(\frac{m}{k_B}\int_{T_0}^{T} \frac{\varepsilon^V_T({T^\prime}  )}{{T^\prime}  ^2} d{T^\prime}   \right),
\end{align*}
where $A^R_0, A^V_0 $ and $T_0$ are inessential constants.
 As is observed in \cite{Wascom2015a,RuggeriSpiga}, the functions $A^R$ and $A^V$ are, according to \eqref{AEi}, the Laplace transforms of $\varphi$ and $\psi$, respectively:
\begin{equation*}
A^R(T) = L_u\left[ \varphi (I^R)\right] (s), \quad A^V(T) = L_u\left[ \pv \right] (s), \quad s=\frac{1}{k_B T},
\end{equation*}
and then we obtain the state functions $\pr$ and $\pv$ as the inverse Laplace transforms of $A^R(T)$ and $A^V(T)$, respectively:
\begin{equation*}
\varphi(I^R)= L_u^{-1}\left[ A^R(T)  \right] (I^R), \  \pv = L_u^{-1}\left[ A^V(T)  \right] (I^V), \  T=\frac{1}{k_B s}.
\end{equation*}

We also notice the relation: 
\begin{align}
p^K(\rho,T) = \frac{2}{3}\rho \varepsilon^K_E(T), \label{pepsi}
\end{align}
and the specific entropy density $s=h_E/\rho$ in equilibrium is given by
\begin{align*}
s = s_E(\rho,T) = s^K_E(\rho, T) + s^R_E(T) + s^V_E(T),
\end{align*}
where
\begin{align}
\begin{split}
s^K_E (\rho,T)&\equiv - \frac{k_B}{\rho} \inta f_E \log f^{(K)}_E \, \pr \pv \da, \\
&= \frac{k_B}{m}\log \left(\frac{{T}^{3/2}}{\rho}\right) + \frac{\varepsilon^K_E(T)}{T} - \frac{k_B}{m} \log \left[\frac{1}{m}\left(\frac{m}{2\pi k_B}\right)^{3/2}\right], \\
s^R_E (T) &\equiv - \frac{k_B}{\rho} \inta f_E \log f^{(R)}_E \, \pr \pv \da, \\
&= \frac{k_B}{m}\log A^R(T) + \frac{\varepsilon^R_E(T)}{T},\\
s^V_E (T) &\equiv - \frac{k_B}{\rho} \inta f_E \log f^{(V)}_E \, \pr \pv \da,\\
&= \frac{k_B}{m}\log A^V(T) + \frac{\varepsilon^V_E(T)}{T}.\\   
\end{split}
\label{sE-rare}
\end{align}
The Gibbs relations of the three modes are given by
\begin{align}
\begin{split}
& T\mathrm{d}s^K_E(\rho, T) = \mathrm{d} \varepsilon^K_E(T) - \frac{p(\rho,T)}{\rho^2}\mathrm{d}\rho, \\
& T\mathrm{d}s^R_E(T) = \mathrm{d}\varepsilon^R_E(T),\quad
 T\mathrm{d}s^V_E(T) = \mathrm{d}\varepsilon^V_E(T).
\end{split}
 \label{Gibbs}
\end{align}

\section{Nonequilibrium triple hierarchy of moment equations}\label{sec:3Noneq}

Before going into a specific ET theory, we briefly make a general discussion on the system of balance equations in ET of polyatomic gases.

Let us introduce three kinds of moments $F$, $H^R$, and $H^V$ as follows:
\begin{align*}
  \begin{split}
 & F_{i_1 \ldots i_j} =  \inta m c_{i_1}\cdots c_{i_j} f \, \pr \pv \da ,   \\
 & H_{lli_1 \ldots i_k}^R =  \inta 2 I^R c_{i_1}\cdots c_{i_k} f \, \pr \pv \da ,  \\
 & H_{lli_1 \ldots i_l}^V  =  \inta 2 I^V c_{i_1}\cdots c_{i_l} f \, \pr \pv \da,   
  \end{split}
\end{align*}
where $j,k,l=1,2, \cdots$.
From the Boltzmann equation \eqref{eq:Boltzmann}, we obtain three hierarchies (a triple hierarchy) of balance equations, i.e., $F$, $H^R$, and $H^V$-hierarchies in the following form:  
\begin{widetext}
\begin{align*}
	 & \partial_t F + \partial_i F_i = 0,  \notag \\
	 & \partial_t F_{i_1} + \partial_i F_{ii_1} = 0, \notag \\
  & \partial_t F_{i_1 i_2} + \partial_i F_{ii_1 i_2} = {P^K_{i_1 i_2}},
   && \partial_t H_{ll}^R + \partial_i H_{lli}^R = {P_{ll}^R},
   && \partial_t H_{ll}^V + \partial_i H_{lli}^V = {P_{ll}^V},  \\
   & \partial_t F_{i_1 i_2 i_3} + \partial_i F_{ii_1 i_2 i_3} = {P^K_{i_1 i_2i_3}},
   && \partial_t H_{lli_1}^R + \partial_i H_{lli i_1}^R = {P_{lli_1}^R},
   && \partial_t H_{lli_1}^V + \partial_i H_{lli i_1}^V = {P_{lli_1}^V}, \notag \\ 
	 & \qquad \ \vdots  && \qquad \ \vdots && \qquad \ \vdots \notag 
\end{align*}
\end{widetext}
where the production terms are related to the collision term as follows:
  \begin{align*}
  \begin{split}
 & P^K_{i_1 \ldots i_j} =  \inta m c_{i_1}\cdots c_{i_j} Q(f) \, \pr \pv \da, \\
 & P_{lli_1 \ldots i_k}^R =   \inta 2I^R  c_{i_1}\cdots c_{i_k} Q(f) \, \pr \pv \da ,\\
 & P_{lli_1 \ldots i_l}^V = \inta 2 I^V c_{i_1}\cdots c_{i_l}  Q(f) \, \pr \pv \da.
  \end{split}
 \end{align*}
 We notice that the first and second equations of the $F$-hierarchy represent the conservation laws of mass and momentum,
 while the sum of the balance equations of $F_{ll}$, $H_{ll}^R$ and $H_{ll}^R$ represents the conservation law of energy with
\begin{align}
 Q_{ll}= P^K_{ll} + P_{ll}^R + P_{ll}^V = 0. \label{P-rare}
\end{align}
In each of the three hierarchies, the flux in one equation appears as the density in the next equation.

  \medskip

  \Remark[\label{R:Gll}] Equivalently, instead of the one of the three hierarchies, we may adopt the hierarchy of the total energy ($G$-hierarchy):
\begin{align*}
&\partial_t G_{ll} + \partial_i G_{lli} = 0,\\
&\partial_t G_{lli_1 \cdots i_m} + \partial_i G_{llii_1\cdots i_m} = Q_{lli_1 \cdots i_m}, \quad m = 1,2, \cdots ,
\end{align*}
where $G_{ll}$ is given by \eqref{hydro}$_3$ and
\begin{align*}
 &G_{lli_1\cdots i_m} = \inta (mc^2+2I^R +2I^V)\\
 &\qquad \qquad \qquad \quad \times c_{i_1} \cdots c_{i_m} f\, \pr \pv \da, 
\end{align*}
and
\begin{align*}
   Q_{lli_1 \cdots i_m} = P^K_{lli_1\cdots i_m}+P^R_{lli_1\cdots i_m} + P^V_{lli_1\cdots i_m} \quad m = 1,2, \cdots.
\end{align*}
The $G$-hierarchy has been introduced in the theory with the binary hierarchy of balance equations (see \eqref{eq:hierarchy-poly}).

\subsection{Truncated system of balance equations and its closure}

To have a finite    system of balance equations, we truncate the $F$, $H^R$, and $H^V$-hierarchies at the orders of $N$, $M$ and $L$, respectively.  For conciseness, it is convenient to introduce a multi-index $A$:
\begin{equation*}
	c_A = \left\{
	\begin{array}{lll}
		1 &  & \text{for } A=0 \\
		c_{i_1} \cdots c_{i_A} &  & \text{for }  A \geq 1
	\end{array}\right.. 
\end{equation*}
The multi-index is also introduced for other quantities in a similar way (see for more details \cite{book}).
Then, we can express the densities as follows:
\begin{align} \label{eq:def_density} 
\begin{split}
 &F_A = \inta  m c_A f \pr \pv \da, \\  
 &H_{llA'}^R = \inta 2 I^R c_{A'} f \pr \pv \da, \\  
 &H_{llA''}^V = \inta 2 I^V c_{A''} f \pr \pv \da.   
\end{split}
\end{align}
The fluxes $F_{iA}$, $H_{lliA'}^R$, $H_{lliA''}^V$ and the productions $P_A^K$, $P_{llA'}^R$, $P_{llA''}^V$ are also expressed in a similar way.

Then a triple hierarchy of moments truncated at the orders $N$, $M$ and $L$ ({$(N,M,L)$-\emph{system}}) is compactly expressed as 
\begin{widetext}
\begin{align} \label{eq:hierarchies-truncated}
	& \partial_t F_A + \partial_i F_{iA} = P_A^K, & & \nonumber \\
	& \quad \left(0 \leq A \leq N\right) & &
 \partial_t H_{llA'}^R + \partial_i H_{lliA'}^R = P_{llA'}^R, && 	\partial_t H_{llA''}^V + \partial_i H_{lliA''}^V = P_{llA''}^V\\
	& & & \qquad \left( 0 \leq A' \leq M\right) && \qquad \left( 0 \leq A'' \leq L\right)\nonumber
\end{align}
\end{widetext}
with $P^K=0$ and $P^K_1=0$ and with the condition \eqref{P-rare} representing the conservation laws of mass, momentum and energy.

\subsubsection{Galilean invariance \label{sec:ETpoly-Galilean}}
Since the velocity-independent variables are the moments in terms of the peculiar velocity $C_i$ instead of $c_i$, it is possible to express the velocity dependence of the densities $\mathbf{F}=(F_A,H_{llA'}^R,H_{llA''}^V)^T$, the non-convective fluxes $\mathbf{\Phi} = (F_{iA}-F_{A}v_i, H_{lliA'}^R - H^R_{llA'} v_i, H_{lliA''}^V - H^V_{llA''}v_i)^T$, and the production terms $\mathbf{P}=(P_{A}^K, P_{llA'}^R, P_{llA''}^V)$ as follows\cite{Ruggeri-1989}:
  \begin{align*}
   \mathbf{F} = \mathbf{X}(\mathbf{v}) \hat{\mathbf{F}}, \quad
   \mathbf{\Phi} = \mathbf{X}(\mathbf{v})  \hat{\mathbf{\Phi}}, \quad
   \mathbf{P} = \mathbf{X}(\mathbf{v})  \hat{\mathbf{P}}, \quad
  \end{align*}
  where a hat on a quantity indicates the velocity-independent part of the quantity.

  We assume that the constitutive quantities $\hat{F}_{iN}$, $\hat{H}^R_{lliM}$, $\hat{H}^V_{lliL}$, $\hat{P}^K_{A}$, $\hat{P}^R_{A'}$, and  $\hat{P}^V_{A''}$, which we express as $\hat{\Psi}$ generically, depend on the densities locally and instantaneously:
 
  \begin{align}
   \hat{\Psi} = \hat{\Psi} (\hat{F}_{A}, \hat{H}^R_{llA'}, \hat{H}^V_{llA''}). \label{eq:closev}
  \end{align}

   \medskip

   \Remark[\label{R:Galilean}] In principle, the truncated orders $N, M$ and $L$ may be chosen independently.  However, if we naturally impose the condition that the $(N,M,L)$-system can make the $G$-hierarchy be Galilean invariant, the inequality; $\min(M,L) \leq N-1$ should be satisfied \cite{Arima-2014} because of the relation:
\begin{align*}
 &G_{lla} = X_{ab}\left(\hat{F}_{llb} + \hat{H}_{llb}^R + \hat{H}_{llb}^V+2v_l \hat{F}_{lb}+v^2 \hat{F}_{b}\right) \\
 &\qquad (0 \leq a,b \leq \min(M,L)).
\end{align*}

\subsubsection{MEP and the closure of the system}

To obtain the constitutive equations \eqref{eq:closev} explicitly, we utilize the MEP.  That is, the most suitable distribution function $f_{(N,M,L)}$ is the one that maximizes the functional defined by  (we omit the symbol of summation for the repeated indices: $A$ from $0$  to $N$, $A'$ from $0$ to $M$, and $A''$ from $0$ to $L$)
\begin{align*}
 &\mathcal{L}_{(N,M,L)}\left(f\right)\\   & = - k_B  \inta f \log f\, \pr \pv \da\\
 &\ \ \ + \lambda_A \left(F_A - \inta m c_A f \pr \pv \da\right)\\
 &\ \ \ + {\mu}_{A'}^R \left( H_{llA'}^R - \inta 2 I^R c_{A'} f \pr \pv \da \right)  \\
 &\ \ \ + {\mu}_{A''}^V \left( H_{llA''}^V - \inta 2 I^V c_{A''} f \pr \pv \da \right),
\end{align*}
where $\lambda_A$, $\mu_{A'}^R$ and ${\mu}_{A''}^V$ are the Lagrange multipliers. As a consequence \cite{Boillat-1997h}, we have
\begin{align*} 
	&f_{(N,M,L)} = \exp \left(-1-\frac{m}{k_B}\chi_{(N,M,L)}\right), \\
	&\chi_{(N,M,L)} = \lambda_A c_A + \frac{2I^R}{m}\mu_{A'}^Rc_{A'} + \frac{2 I^V}{m}\mu_{A''}^Vc_{A''}.
\end{align*} 
Due to the Galilean invariance, the distribution function can be expressed in terms of the velocity-independent quantities:
\begin{align} \label{fNML}
 \begin{split}
	&f_{(N,M,L)} = \exp \left(-1-\frac{m}{k_B}\hat{\chi}_{(N,M,L)}\right), \\
	&\hat{\chi}_{(N,M,L)} = \hat{\lambda}_A C_A + \frac{2I^R}{m}\hat{\mu}^R_{A'}C_{A'} + \frac{2 I^V}{m}\hat{\mu}^V_{A''}C_{A''}.  
 \end{split}
\end{align} 
 Therefore we obtain the velocity dependence of the Lagrange multipliers ${\boldsymbol{\lambda}} \equiv (\lambda_A, \mu^R_{A'}, \mu^V_{A''})$ as follows \cite{Ruggeri-1989,RET}:
\begin{align}
 {\boldsymbol{\lambda}} = \hat{\boldsymbol{\lambda}}\mathbf{X}(-\mathbf{v}). \label{lam-v}
\end{align}

By inserting \eqref{fNML} into $\eqref{eq:def_density}$, the Lagrange multipliers $\lambda_A$, $\mu_{A'}^R$ and $\mu_{A''}^V$ are evaluated in terms of the densities $F_{A}$, $H_{llA'}^R$ and $H_{llA''}^V$. And, finally, by plugging \eqref{fNML} into the last fluxes and production terms, the system is closed.  In this way we obtain the ET theory for the $(N,M,L)$-system. 

\medskip

\Remark{\label{R:EP}} An alternative approach to achieve the closure (phenomenological closure) of the system makes use of the entropy principle. In this case, it is required that all the solutions of \eqref{eq:hierarchies-truncated} satisfy the entropy inequality:
\begin{align*}
	\partial_t h + \partial_i h_i  = \Sigma \geq 0,
\end{align*}
where $h$ is given by \eqref{entropy}, and $h_i$ and $\Sigma$ are the entropy flux and the entropy production defined by
\begin{align} \label{entropyfp}
 \begin{split}
 &h_i = - k_B \inta c_i f \log f \, \pr \pv \da, \quad \\
 & \Sigma = - k_B \inta Q(f) \log f \, \pr \pv \da.  
 \end{split}
\end{align}
According with the general results given first in \cite{Boillat-1997} the two closure methods give the same closed system of balance equations.  Moreover we obtain the following relations:
 \begin{align}
\begin{split}
 &\mathrm{d} h = \lambda_A \mathrm{d}F_A + \mu_{A'}^R \mathrm{d}H_{llA'}^R + \mu_{A''}^V \mathrm{d}H_{llA''}^V,\\
	&\mathrm{d} h_i = \lambda_A \mathrm{d}F_{iA} + \mu_{A'}^R\mathrm{d}H_{lliA'}^R +  \mu_{A''}^V\mathrm{d}H_{lliA''}^V, \\
	&\Sigma = \lambda_A P_A^K + \mu_{A'}^R P_{llA'}^R  + \mu_{A''}^V P_{llA''}^V \geq 0. 
\end{split}
\label{eq:dh}
\end{align}


\section{Generalized BGK-model}\label{sec:prodmol}
 
Concerning the collision term Struchtrup \cite{Struchtrup-1999} and Rahimi and Struchtrup \cite{Struchtrup-2014} proposed a variant of the BGK-model \cite{BGK} to  take into account a relaxation of the energy of the internal mode.
In this section we introduce a novel simple collision term with three relaxation times in order to describe a more refined model in which rotational and vibrational modes are treated individually.

\subsection{Three relaxation times}

\begin{figure}[h]
 \centering
 \includegraphics[width=86mm]{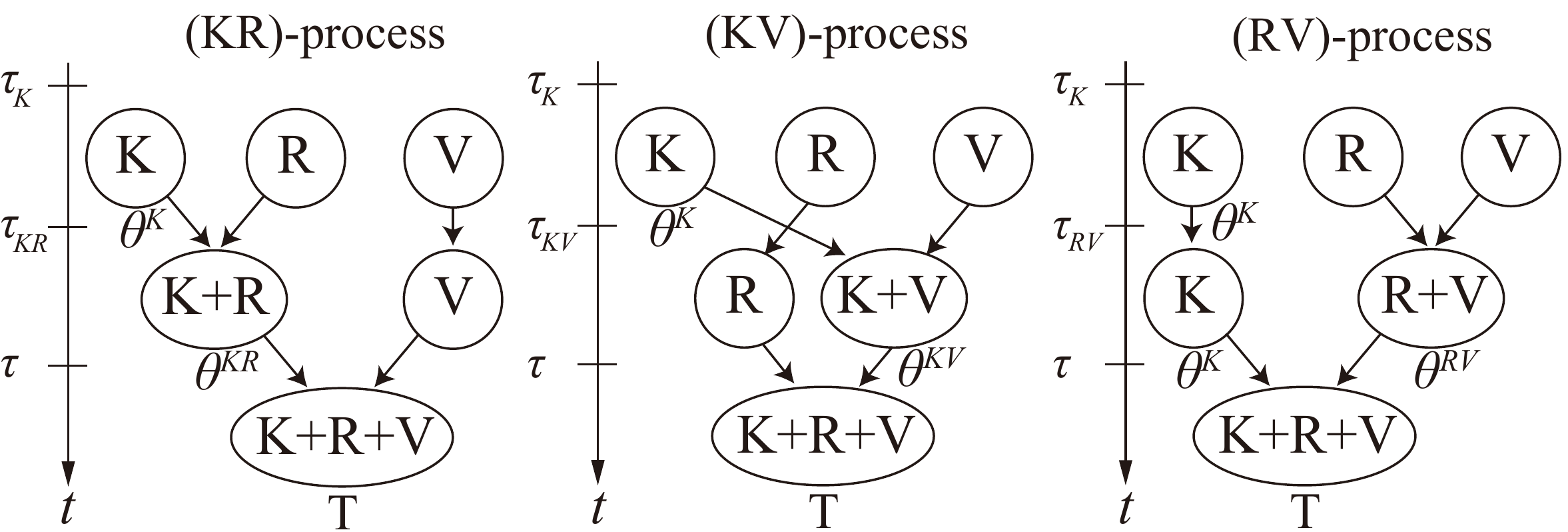} 
 \caption{Diagram of the three possible relaxation processes for the translational mode (K), rotational mode (R), and vibrational mode (V). The symbols $\theta^K$, $\theta^{\mathfrak{bc}}$ (($\mathfrak{b}\mathfrak{c}$) = (KR), (KV), (RV)) are partial equilibrium temperatures and $T$ is the local equilibrium temperature.  A mode without attaching a symbol of the temperature is not necessarily in partial equilibrium.}  
 \label{4process}
\end{figure}

In polyatomic gases, we may introduce three characteristic times corresponding to three relaxation processes caused by the molecular collision (see also \cite{Bhatia,Mason,Stupochenko,Kustova}):
\begin{itemize}
\item[(i)] Relaxation time $\tau_K$: This characterizes the relaxation process within the translational mode (mode K) of molecules.  The process shows the tendency to approach an equilibrium state of the mode K with the distribution function $f_{K:E}$ having the temperature $\theta^K$, explicit expression of which is shown below.  However, the rotational and vibrational modes are, in general, in nonequilibrium.  This process is observable also in monatomic gases.
\item[(ii)] Relaxation time $\tau_{\mathfrak{b}\mathfrak{c}}$: There are energy exchanges among the three modes: mode K, rotational mode (mode R), and vibrational mode (mode V).  The relaxation process occurs in such a way that two of the three modes (say ($\mathfrak{b}\mathfrak{c}$) = (KR), (KV), (RV)) approach, after the relaxation time $\tau_{\mathfrak{b}\mathfrak{c}}$, an equilibrium state characterized by the distribution function $f_{\mathfrak{b}\mathfrak{c}:E}$ with a common temperature $\theta^{\mathfrak{b}\mathfrak{c}}$, explicit expression    of which is shown below. Because of the lack of experimental data, we have no reliable magnitude-relationship between $\tau_K$ and $\tau_{\mathfrak{b}\mathfrak{c}}$.  However it seems natural to adopt the relation: $O(\tau_{\mathfrak{b}\mathfrak{c}}) \gtrsim O(\tau_K)$, which we assume hereafter (see also section \ref{sec:LW}). In Table \ref{table:process}, possible three cases are summarized depending on the choice of $\mathfrak{b}$ and $\mathfrak{c}$.
\item[(iii)] Relaxation time $\tau$ of the last stage: After the relaxation process between $\mathfrak{b}$ and $\mathfrak{c}$, all modes, K, R, and V, eventually approach a local equilibrium state characterized by $f_E$ with a common temperature $T$ among K, R, and V-modes, which is given by \eqref{fE}. Naturally we have the relation: $\tau  > \tau_{\mathfrak{b}\mathfrak{c}}$.
\end{itemize}
Diagrams of the possible relaxation processes are shown in Fig.\ref{4process}. 

\renewcommand{\arraystretch}{1.2}
\begin{table}[h!]
	\centering
	\caption{Three possible relaxation processes in the second stage (ii)}
 \label{table:process}
 \begin{ruledtabular}
	\begin{tabular}{cccc}
	 $(\mathfrak{b}\mathfrak{c})$-Process & $(\mathfrak{a},\mathfrak{b},\mathfrak{c})$ & Relaxation time & Collision term \\ \hline 
	 $(KR)$-process & $(V,K,R)$ & \qquad \quad $\tau_{KR}$ & $Q^{KR}(f)$ \\ 
	  $(KV)$-process & $(R,K,V)$ &\qquad \quad $\tau_{KV}$ & $Q^{KV}(f)$ \\ 
	$(RV)$-process & $(K,R,V)$ &\qquad \quad $\tau_{RV}$ & $Q^{RV}(f)$\\ 
	\end{tabular}
  \end{ruledtabular}
\end{table}
	   \renewcommand{\arraystretch}{1.0}

\subsection{Generalized BGK collision term}

The generalized BGK collision term for $(\mathfrak{b}\mathfrak{c})$-process (($\mathfrak{b}\mathfrak{c}$) = (KR), (KV), (RV)) is proposed as follows:
\begin{align}
Q^{\mathfrak{b}\mathfrak{c}}(f) = - \frac{1}{\tau_K}(f-f_{K:E}) - \frac{1}{\tau_{\mathfrak{b}\mathfrak{c}}}(f-f_{\mathfrak{b}\mathfrak{c}:E}) - \frac{1}{\tau}(f-f_E),
\label{BGK-gen}
\end{align}
where the distribution functions $f_{K:E}$ and $f_{\mathfrak{b}\mathfrak{c}:E}$ are given as follows:

\paragraph{Distribution function $f_{K:E}$:}
This is given by 
\begin{align}
f_{K:E} = \frac{\rho^{RV}(I^R,I^V)}{m }\left(\frac{m}{2\pi k_B \theta^K}\right)^{3/2} \exp \left(-\frac{mC^2}{2 k_B \theta^K}\right), \label{fK}
\end{align}
where
\begin{align}
\rho^{RV} (I^R,I^V) &= \int_{\R^3 } m f d\cc  . \label{rhoi}
\end{align}
This is the equilibrium function with respect to the K-mode with the temperature $\theta^K$ and with the ``frozen'' energies, $I^R$ and $I^V$. In other words, $f_{K:E}$ is a Maxwellian with the mass density $\rho^{RV}(I^R,I^V)$ and temperature $\theta^K$. Therefore  $f_{K:E}$ given in \eqref{fK} is obtained by maximizing not the true entropy \eqref{entropy} but the entropy with the frozen energies:
\begin{equation*} 
h^{RV}(I^R,I^V) = - k_B \int_{\R^3} f \log f\,  d\mathbf{c}
\end{equation*}
under the constraints:
\begin{align*}
\left(\begin{array}{c}
\rho^{RV}(I^R,I^V) \\ \rho^{RV}(I^R,I^V) v_i \\ 2 \rho^{RV}(I^R,I^V) \varepsilon^K_E(\theta^K)
\end{array} \right) =
\int_{\R^3 }
\left(\begin{array}{c}
m\\ m c_i \\ m C^2
\end{array}\right) f
d\cc.
\end{align*}
Then we have the relation \eqref{rhoi}, and the relation:
\begin{align*}
	\varepsilon^{K} = \varepsilon^K_E(\theta^K), \label{}
\end{align*}
from which we can determine the temperature $\theta^K$.

\paragraph{Distribution function $f_{KR:E}$:}

Let us study the process in which the $K$ and $R$-modes reach their common equilibrium with the temperature $\theta^{KR}$ and the vibrational energy $I^V$ can be considered as frozen. In this case, we have the distribution function:
\begin{align*}
f_{KR:E} = \frac{\rho^V(I^V)}{m A^R(\theta^{KR})}\left(\frac{m}{2\pi k_B \theta^{KR}}\right)^{3/2} \exp \left\{-\frac{1}{k_B \theta^{KR}}\left(\frac{mC^2}{2} + I^R\right)\right\},
\end{align*}
where
\begin{align}
\rho^V (I^V) &= m \int_{\R^3 } \int_0^\infty    f \,  \pr \, dI^R d\cc  . \label{rhov}
\end{align}
Since the equilibrium state is described by the mass density $\rho^V (I^V)$ with the frozen vibrational energy $I^V$ and the internal energy $\varepsilon^{K+R}_E (\theta^{KR}) \equiv \varepsilon^K(\theta^{KR}) + \varepsilon^R(\theta^{KR})$, we obtain $f_{KR:E}$ by using the MEP and searching the maximum of the entropy:
\begin{equation*}
h^V(I^V) = -k_B  \int_{\R^3 }  \int_0^\infty f \log f \, \pr  dI^R \, d\cc
\end{equation*}
under the constraints:
\begin{align*}
\left(\begin{array}{c}
\rho^{V}(I^V) \\ \rho^{V}(I^V)  v_i \\ 2 \rho^{V}(I^V)  \varepsilon^{K+R}_E(\theta^{KR})
\end{array} \right) =
 \int_{\R^3 } \int_0^\infty
\left(\begin{array}{c}
m\\ m c_i \\ m C^2 + 2 I^R
\end{array}\right) f \pr  dI^R \, d\cc.
\end{align*}
Therefore we have the relation \eqref{rhov}, and the relation:
\begin{align}
\varepsilon^{K+R} \equiv \varepsilon^{K} + \varepsilon^{R} = \varepsilon^K_E(\theta^{KR}) + \varepsilon^R_E(\theta^{KR}) \equiv \varepsilon^{K+R}_E(\theta^{KR}), \label{TKR}
\end{align}
from which we can determine the temperature $\theta^{KR}$.

\paragraph{Distribution function $f_{KV:E}$:}
In a similar way, we have
\begin{align*}
f^{KV} = \frac{\rho^R(I^R)}{m A^V(\theta^{KV})}\left(\frac{m}{2\pi k_B \theta^{KV}}\right)^{3/2} \exp \left\{-\frac{1}{k_B \theta^{KV}}\left(\frac{mC^2}{2} + I^V\right)\right\},
\end{align*}
where
\begin{align*}
\rho^R (I^R) &= \int_{\R^3 } \int_0^\infty m f \pv d\cc dI^V. 
\end{align*}
And we have the relation, from which we can determine the temperature $\theta^{KV}$:
\begin{align}
\varepsilon^{K+V} \equiv \varepsilon^{K} + \varepsilon^{V} = \varepsilon^K_E(\theta^{KV}) + \varepsilon^V_E(\theta^{KV}) \equiv \varepsilon^{K+V}_E(\theta^{KV}). \label{TKV}
\end{align}

\paragraph{Distribution function $f_{RV:E}$:}
 
In this case the K-mode is in equilibrium with the temperature $\theta^K$, and R and V-modes are also in equilibrium but with the different temperature $\theta^{RV}$.  Then we have the expression similar to \eqref{fE}:
\begin{align*}
f_{RV:E} = \frac{\rho}{m A^R(\theta^{RV})A^V(\theta^{RV})}\left(\frac{m}{2\pi k_B \theta^K}\right)^{3/2} \exp \left(-\frac{mC^2}{2k_B \theta^K} - \frac{I^R+I^V}{k_B \theta^{RV}}\right),
\end{align*}
where the temperature $\theta^{RV}$ is determined by the relation:
\begin{align}
\varepsilon^{R+V} \equiv \varepsilon^{R} + \varepsilon^{V} = \varepsilon^R_E(\theta^{RV}) + \varepsilon^V_E(\theta^{RV}) \equiv \varepsilon^{R+V}_E(\theta^{RV}). \label{TRV}
\end{align}

\paragraph{Distribution function $f_{E}$:} This is the local equilibrium distribution function given by \eqref{fE}, in which the temperature $T$ is given by the condition:
\begin{align}
\varepsilon = \varepsilon_E(T). \label{TT}
\end{align}

\subsection{H-theorem}

From the definition of the distribution functions $f_{K:E}$, $f_{KR:E}$, $f_{KV:E}$, $f_{RV:E}$, and $f_{E}$, it is easy to verify the following relations (($\mathfrak{b}\mathfrak{c}$) = (KR), (KV), (RV)): 
\begin{align*}
 & \inta (f-f_{K:E})\log f_{K:E} \pr \pv \da = 0,\\
 & \inta (f-f_{\mathfrak{b}\mathfrak{c}:E})\log f_{\mathfrak{b}\mathfrak{c}:E} \pr \pv \da = 0,\\
 & \inta (f-f_{E})\log f_E \pr \pv \da = 0.
\end{align*}
Then the entropy production \eqref{entropyfp} can easily be shown to be positive:
\begin{align*}
 \Sigma = k_B &\inta \bigg\{\frac{f-f_{K:E}}{\tau^K}\log \frac{f}{f_{K:E}} + \frac{f-f_{\mathfrak{b}\mathfrak{c}:E}}{\tau_{\mathfrak{b}\mathfrak{c}}}\log \frac{f}{f_{\mathfrak{b}\mathfrak{c}:E}} \\
 & \qquad + \frac{f-f_E}{\tau}\log \frac{f}{f_E} \bigg\}\pr \pv \da
 \geq 0,
\end{align*}
and the H-theorem holds.

\section{ET theory with seven independent fields: ET$_7$}\label{sec:ET7}

The simplest system of \eqref{eq:hierarchies-truncated} next to the Euler system in the present approach is the system with seven independent fields (ET$_7$):
\begin{align}
 &\text{mass density:} && F=\rho,\notag\\
 &\text{momentum density:} && F_i=\rho v_i,\notag\\
 &\text{translational energy density:} && F_{ll}=2\rho \varepsilon^K + \rho v^2, \label{field7}\\
 &\text{rotational energy density:} && H_{ll}^R=2\rho \varepsilon^R ,\notag\\
 &\text{vibrational energy density:} && H_{ll}^V=2\rho \varepsilon^V.\notag
\end{align}
By neglecting the dissipation due to the shear stress and heat flux, the ET$_7$ theory focuses on the description of the internal relaxation processes in a molecule. 

In this section, by means of the kinetic closure, we derive the nonequilibrium distribution function and the closed system of field equations following the general procedure adopted in \cite{ET6,Ruggeri-2016,RuggeriSpiga}.  We assume the generalized BGK-model of the collision term introduced above.

\subsection{System of balance equations}

From \eqref{eq:hierarchies-truncated}, the system of balance equations is expressed as follows:
\begin{align}
	\begin{split}
	 &\frac{\partial F}{\partial t} + \frac{\partial F_i}{\partial x_i} = 0,\\
	 &\frac{\partial F_j}{\partial t} + \frac{\partial F_{ij}}{\partial x_i} = 0,\\
	 &\frac{\partial {F}_{ll}}{\partial t} + \frac{\partial {F}_{lli}}{\partial x_i} = {P}_{ll}^K, \ \
	 \frac{\partial H_{ll}^{R}}{\partial t} + \frac{\partial H_{lli}^{R}}{\partial x_i} = {P}_{ll}^{R}, \  \
	 \frac{\partial H_{ll}^{V}}{\partial t} + \frac{\partial H_{lli}^{V}}{\partial x_i} = {P}_{ll}^{V},
	\end{split}
  \label{ET7}
\end{align}
where ($F_{lli}$, $H_{lli}^R$, $H_{lli}^V$) and ($P_{ll}^K$, ${P}_{ll}^R$, ${P}_{ll}^V$) are the fluxes and productions of the densities ($F_{ll}$, $H_{ll}^R$, $H_{ll}^V$).
It is easily verified that the production terms are velocity independent. 

\subsection{Nonequilibrium distribution function}

First of all, we start with the following statement:

\begin{theorem}
The nonequilibrium distribution function for the truncated system \eqref{ET7} obtained by using the MEP is expressed as
\begin{align}
 f^{(7)} = \frac{\rho}{m A^R(\theta^R)A^V(\theta^V)}\left(\frac{m}{2\pi k_B \theta^K}\right)^{3/2} \exp \left(-\frac{mC^2}{2 k_B \theta^K} - \frac{I^R}{k_B \theta^R} - \frac{I^V}{k_B \theta^V}\right), 
 \label{eq:f7}
\end{align}
 where $A^R(\theta^R)$ and $A^V(\theta^V)$ are normalization factors given by \eqref{AEi}.  Nonequilibrium temperatures $\theta^R$ and $\theta^V$ of R and V-modes are determined through the relations:
 \begin{align*}
  \varepsilon^R = \varepsilon^R_E(\theta^R), \quad \varepsilon^V = \varepsilon^V_E(\theta^V).
 \end{align*}
\end{theorem}

The proof of this statement is given in Appendix.  A similar result was obtained in \cite{RuggeriSpiga} in the case of ET$_6$.
In the present case, as shown also in Appendix, the Lagrange multipliers are given as the functions of $(\rho,v_i,\theta^K,\theta^R,\theta^V)$:
\begin{align}
 \begin{split}
 &\lambda = - \frac{g^K_E(\rho,\theta^K)}{\theta^K} - \frac{g^R_E(\theta^R)}{\theta^R} - \frac{g^V_E(\theta^V)}{\theta^V}+\frac{v^2}{2\theta^K}, \quad \lambda_i = - \frac{v_i}{\theta^K}, \\
 &\mu^K =  \frac{1}{2\theta^K}, \quad \mu^R = \frac{1}{2\theta^R} , \quad \mu^V = \frac{1}{2\theta^V}.   
 \end{split}
 \label{mainfield}
\end{align}
where $g^K_E(\rho,\theta^K), g^R_E(\theta^R)$ and $g^V_E(\theta^V)$ are  the nonequilibrium chemical potentials of the modes:
 \begin{align}
  \begin{split}
 &g^K_E(\rho, \theta^K) = \varepsilon^K_E(\theta^K)+ \frac{p^K(\rho,\theta^K)}{\rho} - \theta^K s^K_E(\rho,\theta^K), \\
 &g^R_E(\theta^R) = \varepsilon^R_E(\theta^R)  - \theta^R s^R_E(\theta^R), \quad
 g^V_E(\theta^V) = \varepsilon^V_E(\theta^V)  - \theta^V s^V_E(\theta^V).   
  \end{split}
  \label{def-g}
 \end{align}

\Remark{\label{R:tauK}} From \eqref{eq:f7}, we notice that, within ET$_7$, any nonequilibrium state can be identified by assigning the nonequilibrium temperatures $\theta^K$, $\theta^R$, and $\theta^V$ together with $\rho$ and $v_i$.  In other words, ET$_7$ adopts the approximation that K, R, and V-modes are always in equilibrium but, in general, with different temperatures from each other.
Therefore ET$_7$ does not take into account the relaxation (i) with the relaxation time $\tau_K$.
See also Fig.\ref{4process}.

\Remark
The nonequilibrium temperatures have been introduced in many studies although there still remain subtle conceptual problems \cite{Jou}. In the context of ET, the nonequilibrium temperature is defined through the Lagrange multiplier corresponding to the conservation law of energy \cite{book,Barbera-1999}. Indeed the expression of the Lagrange multiplies of ET$_7$, \eqref{mainfield}, ensures the present definition of the nonequilibrium temperatures $\theta^K$, $\theta^R$ and $\theta^V$.

\subsection{Closed system of field equations}
 
By using the distribution function \eqref{eq:f7}, we obtain the constitutive equations for the fluxes as follows:
\begin{align}
 \begin{split}
  &F_{ij}  = \inta m c_i c_j f^{(7)} \, \pr \pv \da\\
  &\quad \ = p^K(\rho,\theta^K)\delta_{ij} + \rho v_i v_j ,\\
  &F_{lli}  = \inta m c^2c_i f^{(7)} \, \pr \pv \da\\
  & \quad \  =  \left\{2\rho \varepsilon^{K}_E(\theta^K)+2p^K(\rho, \theta^K) + \rho v^2\right\}v_i,\\
  &H^R_{lli}  = \inta 2 c_i I^R f^{(7)} \, \pr \pv \da \\
  &\quad \ \ =  2\rho \varepsilon^{R}_E(\theta^R)v_i,\\
  &H^V_{lli}  = \inta 2 c_i I^V f^{(7)} \, \pr \pv \da \\
  &\quad \ \ =  2\rho \varepsilon^{V}_E(\theta^V)v_i.  
 \end{split}
 \label{ConstET7}
\end{align}
We notice that the velocity-independent parts of $F_{lli}$, $H^R_{lli}$ and $H^V_{lli}$ vanish.

The trace part of the momentum flux $F_{ll}$ is related to the pressure $p$ and the dynamic pressure $\Pi$ in continuum mechanics as follows:
\begin{align*}
F_{ll} = 3(p+\Pi) + \rho v^2.
\end{align*}
Comparing this relation with \eqref{ConstET7}$_2$, we notice that $\Pi$ is given by
\begin{align}
\Pi = p^K (\rho, \theta^K)-p^K(\rho ,T), \label{Pi2}
\end{align}
or, from \eqref{pepsi}, it is given by
\begin{align*}
\Pi   = \frac{2}{3}\rho \left(\varepsilon^K_E(\theta^K)- \varepsilon^K_E(T)\right). \label{Pi}
\end{align*}
Therefore, as was shown in \cite{denseET,RuggeriSpiga}, the dynamic pressure is related to the energy exchange.

Using the constitutive equations above, we obtain the closed system of field equations for the independent seven fields, $\rho, v_i, \theta^K, \theta^R, \theta^V$ (the equations of state are given by \eqref{ThEqSt} and \eqref{CaEqSt}):
\begin{equation}
\begin{split}
&\frac{\partial \rho}{\partial t}+\frac{\partial}{\partial  x_i}( \rho v_i ) = 0, \\
 &\frac{\partial \rho v_j}{\partial t} + \frac{\partial }{\partial x_i}\left\{ p^K(\rho,\theta^K) \delta_{ij} + \rho v_i v_j\right\} = 0, \\
 &\frac{\partial}{\partial t} \left\{ 2\rho \varepsilon^K_E(\theta^K) +\rho v^2 \right\} + \\
 &+\frac{\partial}{\partial x_i} \left\{ \left(2\rho \varepsilon^K_E(\theta^K)  + \rho v^2 +   2p^K(\rho,\theta^K)\right) v_i\right\} ={P}_{ll}^K ,\\
 &\frac{\partial}{\partial t}\left\{ 2 \rho {\varepsilon}^R_E(\theta^R) \right\} + \frac{\partial}{\partial x_i} \left\{ 2\rho {\varepsilon}^R_E(\theta^R) v_i \right\} =  {P}_{ll}^R ,\\
 &\frac{\partial}{\partial t}\left\{ 2 \rho {\varepsilon}^V_E(\theta^V)\right\} + \frac{\partial}{\partial x_i} \left\{ 2\rho {\varepsilon}^V_E (\theta^V)  v_i \right\} =  {P}_{ll}^V,
\end{split}
\label{balancekc}
\end{equation}
where expressions of the production terms are given in Section \ref{SubSec:Prod}.

By using the material derivative, the system \eqref{balancekc} is rewritten as follows:
\begin{align}
 \begin{split}
 &\dot \rho + \rho  \frac{\partial v_i}{\partial x_i}=0, \\
  &\rho \dot v_i+\frac{\partial p^K(\rho,\theta^K)}{\partial x_i} = 0,\\
  &\dot{\varepsilon}^K_E(\theta^K)  + \frac{p^K(\rho,\theta^K)}{\rho}\frac{\partial v_k}{\partial x_k}=\frac{{P}_{ll}^K}{2\rho},\\
  &\dot{\varepsilon}^R_E(\theta^R) =\frac{{P}_{ll}^R}{2\rho} ,\\
   & \dot{\varepsilon}^V_E(\theta^V) =  \frac{{P}_{ll}^V}{2\rho}.
 \end{split}
 \label{eq:relaxation-Id}  
\end{align}

\subsection{Entropy density and production}

The nonequilibrium specific entropy density $\eta=h/\rho$ for the truncated system \eqref{ET7} is obtained from \eqref{entropy} as follows:
\begin{align}
 \eta = s^K(\rho, \theta^K) + s^R(\theta^R) + s^V(\theta^V), \label{s7}
\end{align}
where $s^K(\rho, \theta^K)$, $s^R(\theta^R)$, and $s^V(\theta^V)$ are calculated by \eqref{sE-rare}.
From \eqref{s7} with \eqref{Gibbs}, we obtain the extension of the Gibbs relation in nonequilibrium as follows:
\begin{align}
 \mathrm{d} \eta= \frac{1}{\theta^K}\left(\mathrm{d}\varepsilon^K - \frac{p^K(\rho,\theta^K)}{\rho^2\theta^K}\mathrm{d} \rho\right) + \frac{1}{\theta^R}\mathrm{d}\varepsilon^R + \frac{1}{\theta^V}\mathrm{d}\varepsilon^V. \label{Gibbs-noneq}
\end{align}

In the present case, the non-convective part of the entropy flux is zero.  Therefore we have
\begin{align*}
 h_i = h v_i.
\end{align*}
Then the balance law of the entropy density is written as follows:
\begin{equation}
\rho \dot{\eta}=\Sigma, \label{eq:entropy-balance}
\end{equation}
where, from \eqref{eq:dh} with \eqref{mainfield}, we obtain the entropy production:
\begin{align*}
 \Sigma 
 & = \frac{{P}_{ll}^K}{2\theta^K} + \frac{{P}_{ll}^R}{2\theta^R} + \frac{{P}_{ll}^V}{2\theta^V} \geq 0.
\end{align*}

\subsection{Production terms in the generalized BGK-model}\label{SubSec:Prod}

As we can prove that $f^{(7)} = f^{(7)}_{K:E}$, the relaxation time $\tau_K$ plays no role in the production term. This is natural from the \textit{Remark \ref{R:tauK}} above. 
By using the collision term for the processes \eqref{BGK-gen}, the production terms are given explicitly as follows:
\begin{itemize}
	\item {Process (KR):}
	\begin{align}
	\begin{split}
	&{P}_{ll}^K  =  - \frac{2\rho}{\tau_{KR}} \left(\varepsilon^K_E(\theta^K) - \varepsilon^K_E(\theta^{KR})\right)
	- \frac{2\rho}{\tau} \left(\varepsilon^K_E(\theta^{K}) - \varepsilon^K_E(T)\right),\\
	&{P}_{ll}^R =   - \frac{2\rho}{\tau_{KR}} \left(\varepsilon^R_E(\theta^R) - \varepsilon^R_E(\theta^{KR})\right)
	- \frac{2\rho}{\tau} \left(\varepsilon^R_E(\theta^{R}) - \varepsilon^R_E(T)\right),\\
	&{P}_{ll}^V = - \frac{2\rho}{\tau} \left(\varepsilon^V_E(\theta^V) - \varepsilon^V_E(T)\right),	
	\end{split}
	\label{Prod-KR}
	\end{align}
	where, from \eqref{TKR}, $\theta^{KR}$ is determined by
	\begin{align*}
	 \varepsilon^{K+R}_E(\theta^{KR}) = \varepsilon^K_E(\theta^{K}) + \varepsilon^R_E(\theta^{R}), 
	\end{align*}
	and, from \eqref{TT}, $T$ is determined by
	\begin{align}
	\varepsilon_E(T) = \varepsilon^K_E(\theta^{K}) + \varepsilon^R_E(\theta^{R}) + \varepsilon^V_E(\theta^{V}). \label{TET7}
	\end{align}
	
	\item {Process (KV):}
	\begin{align*}
	\begin{split}
	&{P}_{ll}^K  =  - \frac{2\rho}{\tau_{KV}} \left(\varepsilon^K_E(\theta^K) - \varepsilon^K_E(\theta^{KV})\right)
	- \frac{2\rho}{\tau} \left(\varepsilon^K_E(\theta^{K}) - \varepsilon^K_E(T)\right),\\
	&{P}_{ll}^R = - \frac{2\rho}{\tau} \left(\varepsilon^R_E(\theta^{R}) - \varepsilon^R_E(T)\right),\\
	&{P}_{ll}^V = - \frac{2\rho}{\tau_{KV}} \left(\varepsilon^V_E(\theta^{V}) - \varepsilon^V_E(\theta^{KV})\right) - \frac{2\rho}{\tau} \left(\varepsilon^V_E(\theta^V) - \varepsilon^V_E(T)\right),	
	\end{split}
	\end{align*}
	where, from \eqref{TKV}, $\theta^{KV}$ is determined by
	\begin{align*}
	\varepsilon^{K+V}_E(\theta^{KV}) = \varepsilon^K_E(\theta^{K}) + \varepsilon^V_E(\theta^{V}), 
	\end{align*}
	and $T$ is determined by \eqref{TET7}.
	
	\item Process (RV):
	\begin{align*}
	\begin{split}
	&{P}_{ll}^K  =  - \frac{2\rho}{\tau} \left(\varepsilon^K_E(\theta^{K}) - \varepsilon^K_E(T)\right) ,\\
	&{P}_{ll}^R = - \frac{2\rho}{\tau_{RV}} \left(\varepsilon^R_E(\theta^R) - \varepsilon^R_E(\theta^{RV})\right)
	- \frac{2\rho}{\tau} \left(\varepsilon^R_E(\theta^{R}) - \varepsilon^R_E(T)\right),\\
	&{P}_{ll}^V = - \frac{2\rho}{\tau_{RV}} \left(\varepsilon^V_E(\theta^V) - \varepsilon^V_E(\theta^{RV})\right)   - \frac{2\rho}{\tau} \left(\varepsilon^V_E(\theta^{V}) - \varepsilon^V_E(T)\right),	
	\end{split}
	\end{align*}
	where, from \eqref{TRV}, $\theta^{RV}$ is determined by
	\begin{align*}
	\varepsilon^{R+V}_E(\theta^{RV}) = \varepsilon^R_E(\theta^{R}) + \varepsilon^V_E(\theta^{V}), 
	\end{align*}
	and $T$ is determined by \eqref{TET7}.

\end{itemize}

\section{Characteristic features of ET$_7$}\label{sec:FeaturesET7}

We summarize some features of the ET$_7$ theory.

\subsection{Comparison with the Meixner theory}\label{subsec:Meixner}

Thermodynamic theories with internal variables have been developed \cite{de Groot,Coleman,Maugin1,Maugin2,Van2008}, the prototype of which is the Meixner theory  \cite{Meixner-1943,Meixner-1952,de Groot}.
The system of field equations of the Meixner theory with two internal variable $\xi^{(1)}$ and $\xi^{(2)}$ is expressed as follows: 
\begin{equation}\label{eq:Meixner}
\begin{split}
&\dot \rho + \rho\frac{\partial  v_i}j{\partial  x_i} = 0,\\
&\rho \dot v_i+\frac{\partial  \mathcal{P}}{\partial  x_i}
  = 0,\\
&\rho  \dot{\mathcal{E}}
  + \mathcal{P} \frac{\partial  v_k}{\partial  x_k}=0, \\
 &\dot \xi^{(1)} = - \beta^{(1)} \mathcal{A}^{(1)}, \\
 &\dot \xi^{(2)} = - \beta^{(2)} \mathcal{A}^{(2)}, 
\end{split}
\end{equation}
where $\mathcal{P}$, $\mathcal{E}$ and $\mathcal{A}^{(a)}(a=1,2)$ are, respectively, the pressure, the specific internal energy and the affinities of the relaxation processes, 
and $\beta^{(a)}$ are positive phenomenological coefficients. The generalized Gibbs relation in the Meixner theory is assumed to be: 
\begin{equation}\label{eq:GibbsM}
\mathcal{T} {\rm d}\mathcal{S} = {\rm d}\mathcal{E} - \frac{\mathcal{P}}{\rho^2}{\rm d}\rho - \sum_{a=1}^{2}\mathcal{A}^{(a)} {\rm d}\xi^{(a)}, 
\end{equation}
where $\mathcal{T}$ is the temperature and $\mathcal{S}$ is the specific entropy.  Note that the quantities $\mathcal{T}$, 
$\mathcal{S}$, $\mathcal{P}$ and $\mathcal{A}$ 
depend not only on the mass density $\rho$ and the specific internal energy $\mathcal{E}$ but also on the internal variables $\xi^{(a)}$.
From \eqref {eq:GibbsM}, with the use of \eqref {eq:Meixner}, we obtain
\begin{equation}
\dot{\mathcal{S}}= \frac{1}{\mathcal{T}} \sum_{a=1}^{2}\beta^{(a)} {\mathcal{A}^{(a)}}^2. \label{Mentropy}
\end{equation}

Comparing the system of the ET$_7$ theory \eqref{eq:relaxation-Id}, \eqref{Gibbs-noneq} and \eqref{eq:entropy-balance} with the system of the Meixner theory \eqref{eq:Meixner}, \eqref{eq:GibbsM} and \eqref {Mentropy}, we have the following relationship between the Meixner theory and the ET$_7$ theory:
\begin{equation*}
\begin{split}
&\xi^{(1)} = \varepsilon^R_E(\theta^R), \ \  \xi^{(2)} = \varepsilon^V_E(\theta^V),\\
 &\mathcal{P} = p^K(\rho,\theta^K),  \ \ 
 \mathcal{E} = \varepsilon_E(T),  \ \  \mathcal{S} = \eta(\rho,\theta^K,\theta^R,\theta^V),  \ \  \mathcal{T}=\theta^K,\\
&\mathcal{A}^{(1)} = -  \theta^K \left(\frac{1}{\theta^R} - \frac{1}{\theta^K}\right), \ \ \mathcal{A}^{(2)} = -  \theta^K \left(\frac{1}{\theta^V} - \frac{1}{\theta^K}\right), \\
 &\beta^{(1)} = \frac{1}{2\rho \theta^K}\left(\frac{1}{\theta^R} - \frac{1}{\theta^K}\right)^{-1}P_{ll}^R, \ \ 
 \beta^{(2)} = \frac{1}{2\rho \theta^K}\left(\frac{1}{\theta^V} - \frac{1}{\theta^K}\right)^{-1}P_{ll}^V. 
\end{split}
\end{equation*}

To sum up, we have identified the quantities in the Meixner theory in terms of the more understandable quantities of ET$_7$.
In particular, the nonequilibrium temperature $\mathcal{T}$ is recognized as the temperature of the translational mode $\theta^K$. This is reasonable because, from \eqref{eq:GibbsM}, $\mathcal{T}(=\theta^K)$ is the temperature of a state in equilibrium under a constraint that the system is kept at fixed values of $\xi^{(a)}(=\varepsilon^R_E(\theta^R), \varepsilon^V_E(\theta^V))$.

\subsection{Characteristic velocity, sub-characteristic conditions, and local exceptionality}

It is well known that the characteristic velocity $V$ associated with a hyperbolic system of equations can be obtained by using the operator chain rule (see \cite{book}):
\begin{equation*}
\frac{\partial }{\partial t} \, \rightarrow - V \delta, \quad \frac{\partial }{\partial x_i} \, \rightarrow n_i\delta, \quad \mathbf{f}  \rightarrow 0,
\end{equation*} 
 where $n_i$ denotes the $i$-component of the unit normal to the wave front, $\mathbf{f}$ is the production terms and $\delta$ is a differential operator \cite{book}.  In the present case, 
if we choose $\{\rho, v_i, \eta, \theta^R,\theta^V\}$ as independent variables instead of $\{\rho, v_i, \theta^K,\theta^R,\theta^V\}$, and adopt the entropy law \eqref{eq:entropy-balance} instead of the energy equation of the $K$-mode in \eqref{eq:relaxation-Id}$_3$, we obtain 
\begin{align}
&\textbf{Contact Waves:} \quad V= v_n =0, \label{contact} \\
& \ \ \text{(multiplicity 5)} \nonumber\\
& \textbf{Sound Waves:} \quad V =v_n  \pm  \sqrt{\left(\frac{\partial p(\rho,\theta^{K}(\rho,\eta,\theta^R,\theta^V))}{\partial \rho}\right)_{\eta,\theta^R,\theta^V}}\,\, \label{sonic}\\
& \ \ \text{(each of multiplicity 1)}, \nonumber
\end{align}
where $ v_n= v_j n_j$. Here and hereafter, $p^K$ is denoted by $p$ for simplicity.
We can rewrite the velocity of the sound wave $U= V-v_n$ as follows:
\begin{align*}
 U^2 
&= p_{\rho}(\rho,\theta^{K}) + \frac{\theta^{K} p^2_{\theta^{K}}(\rho,\theta^{K})}{\rho^2 c_v^{K}(\theta^{K})},
\end{align*}
where a subscript attached to $p$  indicates a partial derivative and $c_v^K$ is the specific heat of the translational mode defined by $c_v^K(T)=\mathrm{d} \varepsilon^K_E(T)/\mathrm{d} T$.
In an equilibrium case, we have
\begin{align*}
 U^2_E = p_{\rho}(\rho,T) + \frac{T p^2_{T}(\rho,T)}{\rho^2 c_v^{K}(T)}.
\end{align*}

The sound velocity of the Euler fluid is given by
\begin{align*}
U^2_{Euler} = p_{\rho}(\rho,T) + \frac{T p^2_{T}(\rho,T)}{\rho^2 c_v(T)},
\end{align*}
where $c_v$ is the specific heat defined by $c_v(T)=\mathrm{d} \varepsilon_E (T)/\mathrm{d} T$ and 
\begin{align*}
 c_v = c_v^K + c_v^R + c_v^V
\end{align*}
with  the specific heat of the rotational mode $c_v^R$ and the vibrational mode $c_v^V$: $c_v^R(T)=\mathrm{d} \varepsilon^R_E(T)/\mathrm{d} T$ and $c_v^V(T)=\mathrm{d} \varepsilon^V_E(T)/\mathrm{d} T$.
Since the specific heats of the three modes are positive, we notice that the subcharacteristic condition \cite{Boillat-1997h} is satisfied:
 \begin{align*}
   U_E >  U_{Euler}.
 \end{align*}

It is well known that a characteristic velocity associated with a wave is classified as (see e.g. \cite{book}):
\emph{genuinely non-linear} if  
$\delta V =  \nabla_\mathbf{u}V \cdot \delta \mathbf{u} \,\, \propto \,\,
 \nabla_\mathbf{u}V \cdot \mathbf{r} \neq 0, \,\,\,   \forall \mathbf{u}$;
 \emph{linearly degenerate or exceptional} if  
 $\delta V  \equiv 0, \,\,\,   \forall \mathbf{u}$;
 \emph{locally linearly degenerate or locally exceptional} if  
 $  \delta V  = 0, \,\,\,   \text{for some} \,\,\, \mathbf{u},$
where $\mathbf{r}$ is the corresponding eigenvector associated to the system \eqref{ET7}.
The contact waves \eqref{contact} are exceptional while the sound waves \eqref{sonic} can be locally exceptional if the condition is satisfied. Simple algebra similar to the one in \cite{Zhao} gives that, if the hyper-surface of local exceptionality exists, the following relation is satisfied on it:
\begin{align*}
 \begin{split}
  \delta V = \frac{1}{2\rho^2 U} \left(\frac{\partial \rho^2  U^2}{\partial \rho } \right)_{\eta,\theta^R,\theta^V} = 0. 
 \end{split}
\end{align*}

The results obtained here will be useful in the analysis of nonlinear waves such as shock waves
.

\subsection{ET$_6$ theories as the principal subsystems of the ET$_7$ theory}\label{sec:subsystem}

Let us consider the $(\mathfrak{b}\mathfrak{c})$-process (($\mathfrak{b}\mathfrak{c}$) = (KR), (KV), (RV)) defined in  \eqref{BGK-gen} again, and assume that the relaxation time $\tau$ is of several orders larger than the relaxation time $\tau_{\mathfrak{b}\mathfrak{c}}$.  In such a case, the composite system of $\mathfrak{b}$-mode and $\mathfrak{c}$-mode quickly reaches a state with the common temperature $\theta^{\mathfrak{b}\mathfrak{c}}$. Therefore, except for the short period of $O(\tau_{\mathfrak{b}\mathfrak{c}})$ after the initial time, we have the relation:
\begin{align*}
\theta^\mathfrak{b} = \theta^\mathfrak{c} = \theta^{\mathfrak{b}\mathfrak{c}}.
\end{align*}

 As the balance equation of the density  $\left(\varepsilon^\mathfrak{b}_E(\theta^\mathfrak{b})-\varepsilon^\mathfrak{c}_E(\theta^\mathfrak{c})\right)$ is identically satisfied in the present approximation, the remaining equations are given by 
\begin{equation*}
\begin{split}
&\frac{\partial \rho}{\partial t}+\frac{\partial}{\partial  x_i}( \rho v_i ) = 0, \\
&\frac{\partial \rho v_j}{\partial t} + \frac{\partial }{\partial x_i}\left\{ (p+\Pi)\delta_{ij} + \rho v_i v_j\right\} = 0, \\
&\frac{\partial}{\partial t} \left\{ 2\rho \varepsilon +\rho v^2 \right\} + \frac{\partial}{\partial x_i} \left\lbrace   \left(2\rho \varepsilon +   2 p + 2\Pi + \rho v^2\right)v_i\right\rbrace  = 0,\\
&\frac{\partial (2\rho \mathcal{E})}{\partial t} + \frac{\partial (2\rho \mathcal{E}_i)}{\partial x_i} =  P_{\mathcal{E}},
\end{split}
\end{equation*}
where $\mathcal{E}$ is the nonequilibrium energy density characterizing the relaxation process, and $\mathcal{E}_i$ and $P_{\mathcal{E}}$ are its flux and production.  We may regard this system as the ET theory with $6$ fields, which we call ET$_6^{\mathfrak{b}\mathfrak{c}}$.
In Table \ref{table:abc}, three possible ET$_6$ theories corresponding the types of the relaxation process are summarized.  
\renewcommand{\arraystretch}{1.2}
	   \begin{table}[h!]
		\centering
		\caption{Three possible ET$_6$ theories.}
		\label{table:abc}
  \begin{ruledtabular}
		 \begin{tabular}{ccccccc}
		 &Process & $(\mathfrak{a},\mathfrak{b},\mathfrak{c})$ & $p+\Pi$& $\mathcal{E}$ &$\mathcal{E}_i$ & $P_\mathcal{E}$ \\ \hline 
		 ET$_6^{KR}$&$(KR)$ & $(V,K,R)$ & $p(\rho, \theta^{KR})$&  $ \varepsilon^V_E(\theta^V)$ &  $ \varepsilon^V_E(\theta^V)v_i$ & $P_{ll}^V$ \\ 
		 ET$_6^{KV}$& $(KV)$ & $(R,K,V)$ & $p(\rho, \theta^{KV})$& $\varepsilon^R_E(\theta^R)$ &$\varepsilon^R_E(\theta^R)v_i$ & $P_{ll}^R$\\ 
		 ET$_6^{RV}$&$(RV)$ & $(K,R,V)$ & $p(\rho, \theta^{K})$& $ \varepsilon^{RV}_E(\theta^{RV})$ &$ \varepsilon^{RV}_E(\theta^{RV})v_i$ & $P_{ll}^R + P_{ll}^V$\\ 
		 \end{tabular}
   \end{ruledtabular}
	   \end{table}
\renewcommand{\arraystretch}{1.0}

The above argument can be rigorously formulated by using the idea of the principal subsystem \cite{RET}.  In the present case, ET$_6$ is the principal subsystem of ET$_7$. The crucial point is that, all the universal principles of continuum thermomechanics -- objectivity, entropy, and causality principles -- are automatically preserved also in the subsystem.

The characteristic velocity of ET$_6^{RV}$ is obtained as
\begin{align*}
	 {U^{RV}}^2 
	 &= p_{\rho}(\rho,\theta^{K}) + \frac{\theta^{K} p^2_{\theta^{K}}(\rho,\theta^{K})}{\rho^2 c_v^{K}(\theta^{K})},
\end{align*}
which is the same as the characteristic velocity of ET$_7$: $U^{RV} = U$. On the other hand, for ET$_6^{\mathfrak{b}\mathfrak{c}}$ with $(\mathfrak{b},\mathfrak{c}) = (K,R)$ or $(K,V)$, we obtain
\begin{align*}
  {U^{\mathfrak{b}\mathfrak{c}}}^2 
  &= p_{\rho}(\rho,\theta^{\mathfrak{b}\mathfrak{c}}) + \frac{\theta^{\mathfrak{b}\mathfrak{c}} p^2_{\theta^{\mathfrak{b}\mathfrak{c}}}(\rho,\theta^{\mathfrak{b}\mathfrak{c}})}{\rho^2 c_v^{\mathfrak{b}+\mathfrak{c}}(\theta^{\mathfrak{b}\mathfrak{c}})},
\end{align*}
where $c_v^{\mathfrak{b}+\mathfrak{c}} = c_v^\mathfrak{b} + c_v^\mathfrak{c}$.  
Since $c_v > c_v^{\mathfrak{b}+\mathfrak{c}} > c_v^K$, we have the following relation in equilibrium:
\begin{align*}
 U_E > U_E^{\mathfrak{b}\mathfrak{c}} > U_{Euler}.
\end{align*}

\Remark
The ET$_6$ theory studied in the previous papers \cite{ET6,ET6Meccanica,ET6shock,ET6nonlinear,ET6nonlinearshock,Wascom2015a,Wascom2015b} directly corresponds to the ET$_6^{RV}$ theory in the present notation.  However, it should be noted that the previous ET$_6$ theory may also correspond to the ET$_7$ theories with (KR) and (KV)-processes as far as the V-mode is kept in the ground state and has no role in the phenomena under study.

\subsection{Near equilibrium case}

In the $(\mathfrak{b}\mathfrak{c})$-process (($\mathfrak{b}\mathfrak{c}$) = (KR), (KV), (RV)), energy exchanges among $\mathfrak{a}$, $\mathfrak{b}$ and $\mathfrak{c}$-modes are characterized by the following quantities:
 \begin{align*}
  &\delta \equiv \varepsilon^\mathfrak{b}_E(\theta^\mathfrak{b})-\varepsilon^\mathfrak{b}_E(\theta^{\mathfrak{b}\mathfrak{c}}) =  - \varepsilon^\mathfrak{c}_E(\theta^\mathfrak{c}) + \varepsilon^\mathfrak{c}_E(\theta^{\mathfrak{b}\mathfrak{c}}), \\
  &\Delta \equiv \varepsilon^\mathfrak{a}_E(\theta^\mathfrak{a}) - \varepsilon^\mathfrak{a}_E(T) = -\varepsilon^{\mathfrak{b}+\mathfrak{c}}_E(\theta^{\mathfrak{b}\mathfrak{c}}) + \varepsilon^{\mathfrak{b}+\mathfrak{c}}_E(T).
 \end{align*}
By expanding the nonequilibrium energies of the three modes with respect to the nonequilibrium temperatures around an equilibrium temperature $T$ up to the first order, we obtain
\begin{align*}
 &\delta = c_v^\mathfrak{b}(\theta^\mathfrak{b} - \theta^{\mathfrak{b}\mathfrak{c}}) = - c_v^\mathfrak{c}(\theta^\mathfrak{c} - \theta^{\mathfrak{b}\mathfrak{c}}) \\
 &\Delta = c_v^\mathfrak{a} (\theta^\mathfrak{a} - T) = -c_v^{\mathfrak{b}+\mathfrak{c}}(\theta^{\mathfrak{b}\mathfrak{c}}-T).
\end{align*}
Here and hereafter we use the notation $c_v^\mathfrak{a}$ instead of $c_v^\mathfrak{a}(T)$ and so on for simplicity.
Inversely, the nonequilibrium temperatures are expressed as follows:
\begin{align*}
 &\theta^\mathfrak{a}-T = \frac{\Delta}{c_v^\mathfrak{a}}, \quad  \theta^{\mathfrak{b}\mathfrak{c}}-T = - \frac{\Delta}{c_v^{\mathfrak{b}+\mathfrak{c}}}, \\
 &\theta^\mathfrak{b}-T = \frac{\delta}{c_v^\mathfrak{b}} - \frac{\Delta}{c_v^{\mathfrak{b}+\mathfrak{c}}},  \quad 
 \theta^\mathfrak{c}-T = - \frac{\delta}{c_v^\mathfrak{c}} -  \frac{\Delta}{c_v^{\mathfrak{b}+\mathfrak{c}}}.
\end{align*}

The production terms are now given by
\begin{align*}
   \begin{split}
	 &{P}_{ll}^\mathfrak{a} = -2\rho \frac{\Delta}{\tau}, \\
	&{P}_{ll}^\mathfrak{b} = -2\rho \frac{\delta}{\tau_\delta} + 2\rho \frac{c_v^\mathfrak{b}}{c_v^{\mathfrak{b}+\mathfrak{c}}}\frac{\Delta}{\tau},\\
	 &{P}_{ll}^\mathfrak{c} = 2\rho \frac{\delta}{\tau_\delta} + 2\rho\frac{c_v^\mathfrak{c}}{c_v^{\mathfrak{b}+\mathfrak{c}}}\frac{\Delta}{\tau},
   \end{split}
\end{align*}
where $\tau_\delta$ is defined as
\begin{align*}
 \frac{1}{\tau_\delta} \equiv \frac{1}{\tau_{\mathfrak{b}\mathfrak{c}}} + \frac{1}{\tau}.
\end{align*}
Then the entropy production is given by
\begin{align*}
 \Sigma = \frac{\rho}{T^2 } \frac{c_v^{\mathfrak{b}+\mathfrak{c}}}{c_v^\mathfrak{b} c_v^\mathfrak{c}}\frac{1}{\tau_\delta}\delta^2 +  \frac{\rho}{T^2 } \frac{c_v}{c_v^{\mathfrak{a}} c_v^{\mathfrak{b}+\mathfrak{c}}}\frac{1}{\tau}\Delta^2 .
\end{align*}
Since $c_v^K >0, \ c_v^R>0$ and $c_v^V>0$, and $\tau_{\delta} >0, \ \tau > 0$, the entropy production is non-negative.

The system of field equations \eqref{eq:relaxation-Id} is rewritten as follows:
\begin{align}
 \begin{split}
 &\dot \rho + \rho  \frac{\partial v_i}{\partial x_i}=0, \\
  &\rho \dot v_i+\frac{\partial }{\partial x_i}(p + \Pi ) = 0,\\
  &\rho c_v \dot T + (p + \Pi)\frac{\partial v_i}{\partial x_i}=0,\\
   & \dot{\delta}  + \frac{p + \Pi}{\rho} \left\{A_1 + \frac{1}{c_v}\frac{\mathrm{d} }{\mathrm{d}T}\left(\frac{c_v^\mathfrak{b}}{c_v^{\mathfrak{b}+\mathfrak{c}}}\right)\Delta\right\}\frac{\partial v_i}{\partial x_i} = - \frac{\delta}{\tau_{\delta}},\\
 & \dot{\Delta}  + \frac{p + \Pi}{\rho}\frac{A_2-c_v^\mathfrak{a}}{c_v} \frac{\partial v_i}{\partial x_i}=  -\frac{\Delta}{\tau},
 \end{split}
 \label{relax2}  
\end{align}
where $p = p(\rho, T)$, and $A_1$, $A_2$ and $\Pi$ are given in Table \ref{Table:A}. In the limit $\tau_\delta \to 0$, this system reduces to the system of ET$_6^{\mathfrak{b}\mathfrak{c}}$.
 \begin{table}[h!] 
  \centering
  \caption{Explicit expression of $A_1$, $A_2$ and $\Pi$}
  \begin{ruledtabular}
  \begin{tabular}{cccc}
  $(\mathfrak{b}\mathfrak{c})$ & $A_1$ & $A_2$ & $\Pi$\\ \hline 
  $(KR)$ or $(KV)$ & $\displaystyle \frac{c_v^\mathfrak{c}}{c_v^{\mathfrak{b}+\mathfrak{c}}}$ &0 & $\displaystyle \frac{p_T}{c_v^\mathfrak{b}}\delta - \frac{p_T}{c_v^{\mathfrak{b}+\mathfrak{c}}}\Delta$\\ 
  $(RV)$ & $0$& $c_v$ & $\displaystyle \frac{p_T}{c_v^{K}}\Delta$\\
  \end{tabular}   
  \end{ruledtabular}
  \label{Table:A}
 \end{table}


When we apply the Maxwellian iteration \cite{Ikenberry} on \eqref{relax2}$_{4,5}$ and retain the first order terms with respect to the relaxation times $\tau_{\delta}$ and $\tau$, we obtain the following approximations for small relaxation times:
\begin{align*}
 \delta =- \tau_{\delta} \frac{p}{\rho} A_1 \frac{\partial v_i}{\partial x_i}, \qquad
 \Delta = -\tau \frac{p}{\rho} \frac{A_2-c_v^\mathfrak{a}}{c_v}\frac{\partial v_i}{\partial x_i}.
\end{align*}

For $(\mathfrak{b}\mathfrak{c})$-process ($(\mathfrak{b}\mathfrak{c})=$(KR) or (KV)), from \eqref{Pi2}, we have
\begin{align*}
 \Pi^{\mathfrak{b}\mathfrak{c}} = -\tau_{\delta} p \frac{\hat{c}_v^\mathfrak{c}}{\hat{c}_v^\mathfrak{b}\hat{c}_v^{\mathfrak{b}+\mathfrak{c}}}\frac{\partial v_i}{\partial x_i} \qquad
  \Pi^{\mathfrak{a}} = -\tau p \frac{\hat{c}_v^\mathfrak{a}}{\hat{c}_v^{\mathfrak{b}+\mathfrak{c}}\hat{c}_v}\frac{\partial v_i}{\partial x_i},
\end{align*}
where $\hat{c}_v = c_v/(k_B/m)$, $\hat{c}_v^K = c_v^K/(k_B/m)$, $\hat{c}_v^{K+R} = c_v^{K+R}/(k_B/m)$, and $\hat{c}_v^V = c_v^V/(k_B/m)$.
Recalling the definition of the bulk viscosity $\nu$:
\begin{align*}
\Pi = -\nu \frac{\partial v_i}{\partial x_i}, 
\end{align*}
we have its expression as follows:
\begin{align*}
 \nu = \tau_{\delta} p \frac{\hat{c}_v^\mathfrak{c}}{\hat{c}_v^\mathfrak{b}\hat{c}_v^{\mathfrak{b}+\mathfrak{c}}} + \tau p \frac{\hat{c}_v^\mathfrak{a}}{\hat{c}_v^{\mathfrak{b}+\mathfrak{c}}\hat{c}_v} .
\end{align*}
This expression is a generalization of the previous results \cite{ChapmanCowling,Tisza}.
If $\tau >> \tau_{\delta}$ and $\hat{c}_v^\mathfrak{a}$ has a value of $O(1)$, the bulk viscosity is approximated by 
\begin{align}
 \nu^{\mathfrak{a}} = \tau p \frac{\hat{c}_v^\mathfrak{a}}{\hat{c}_v^{\mathfrak{b}+\mathfrak{c}}\hat{c}_v}
 = (({\tilde{U}^{\mathfrak{b}+\mathfrak{c}}}_E)^2 - (\tilde{U}_{Euler})^2 )  {\tau}p,
 \label{nuv}
\end{align}
where $\tilde{U}^{\mathfrak{b}+\mathfrak{c}}_E = {U^{\mathfrak{b}+\mathfrak{c}}_E} / \sqrt{k_B T/m}$ and  $\tilde{U}_{Euler} = {U_{Euler}} / \sqrt{k_B T/m}$.  This expression can be derived also from ET$_6^{\mathfrak{b}\mathfrak{c}}$.

For (RV)-process, we have
\begin{align*}
 \delta=0, \qquad \Pi = -\tau p \frac{\hat{c}_v^{R+V}}{\hat{c}_v^K\hat{c}_v}\frac{\partial v_i}{\partial x_i}, 
\end{align*}
and the bulk viscosity is evaluated as
\begin{align*}
 \nu = \tau p \frac{\hat{c}_v^{R+V}}{\hat{c}_v^K\hat{c}_v},
\end{align*}
which is the same as the one derived from ET$_6^{RV}$ \cite{ET6,ET6Meccanica}.

\subsection{Homogeneous solution and relaxation of nonequilibrium temperatures}\label{}

In order to focus our attention on the behavior of the internal molecular relaxation processes, we study first a simple case: homogeneous solutions of the system \eqref{eq:relaxation-Id}, i.e., solutions in which the unknowns are independent of space coordinates and depend only on the time $t$. The system \eqref{eq:relaxation-Id} reduces now to an ODE system:
\begin{align}
 \begin{split}
 &\frac{d \rho}{dt}   =0, \\
  &\frac{d \mathbf{v}}{dt} = 0,\\
  &\frac{d{\varepsilon}^K_E(\theta^K)}{dt}  =\frac{{P}_{ll}^K}{2\rho},\\
  &\frac{d{\varepsilon}^R_E(\theta^R)}{dt} =\frac{{P}_{ll}^R}{2\rho} ,\\
   & \frac{d{\varepsilon}^V_E(\theta^V) }{dt}=  \frac{{P}_{ll}^V}{2\rho}.
 \end{split}
 \label{eq:relaxation-ode}  
\end{align}
The first two equations give that  $\rho$ and $\mathbf{v}$ are constant and for Galilean invariance we can assume without any loss of generality that $\mathbf{v}=0$. Moreover, from \eqref{TET7}, summing the last three equations of \eqref{eq:relaxation-ode}, and taking into account that the sum of the productions is zero and that $\varepsilon_E(T)$ is monotonous function,  we conclude that also $T$ is constant.  Therefore there remain only the last three equations of \eqref{eq:relaxation-ode} that govern the relaxation of the nonequilibrium temperatures. 

For simplicity, we now assume a process near equilibrium and then consider a linearized version. Taking into account \eqref{Prod-KR}, we obtain the following linear ODE system:   
\begin{align}
 \begin{split}
   &\frac{d \bar{\theta}^\mathfrak{a}}{dt} = - \frac{1}{\tau}\bar{\theta}^\mathfrak{a} ,\\
 &\frac{d \bar{\theta}^\mathfrak{b}}{dt} = - \frac{1}{\tau}\bar{\theta}^\mathfrak{b} - \frac{1}{\tau_{\mathfrak{b}\mathfrak{c}}}(\bar{\theta}^\mathfrak{b} - \bar{\theta}^{\mathfrak{b}\mathfrak{c}}),\\
 &\frac{d \bar{\theta}^\mathfrak{c}}{dt}  = - \frac{1}{\tau}\bar{\theta}^\mathfrak{c} - \frac{1}{\tau_{\mathfrak{b}\mathfrak{c}}}(\bar{\theta}^\mathfrak{c} - \bar{\theta}^{\mathfrak{b}\mathfrak{c}}),
 \end{split}
\label{gasatrest}
\end{align}
where $\bar{\theta}^\mathfrak{a} \equiv \theta^\mathfrak{a} -T$, $\bar{\theta}^\mathfrak{b}\equiv \theta^\mathfrak{b} - T$, $\bar{\theta}^\mathfrak{c}\equiv \theta^\mathfrak{c} - T$, and
\begin{align}
 \bar{\theta}^{\mathfrak{b}\mathfrak{c}} \equiv \theta^{\mathfrak{b}\mathfrak{c}}-T = \frac{c_v^\mathfrak{b}\bar{\theta}^\mathfrak{b} + c_v^\mathfrak{c} \bar{\theta}^\mathfrak{c}}{c_v^{\mathfrak{b}+\mathfrak{c}}}.
 \label{tbarKR}
\end{align}
The solution with the initial data $\bar{\theta}_0^\mathfrak{a} = \bar{\theta}^\mathfrak{a}|_{t=0}$, $\bar{\theta}_0^\mathfrak{b} = \bar{\theta}^\mathfrak{b} |_{t=0}$ and $\bar{\theta}_0^\mathfrak{c} = \bar{\theta}^\mathfrak{c}|_{t=0}$  is given by
\begin{align}
 \begin{split}
   &\bar{\theta}^\mathfrak{a} = \bar{\theta}_0^\mathfrak{a}  \mathrm{e}^{-\hat{t}},\\
 &\bar{\theta}^\mathfrak{b} = \frac{1}{\hat{c}_v^{\mathfrak{b}+\mathfrak{c}}}( \hat{c}_v^\mathfrak{b} \bar{\theta}_0^\mathfrak{b} + \hat{c}_v^\mathfrak{c} \bar{\theta}_0^\mathfrak{c} )\mathrm{e}^{-\hat{t}} + \frac{\hat{c}_v^\mathfrak{c}}{\hat{c}_v^{\mathfrak{b}+\mathfrak{c}}}(\bar{\theta}_0^\mathfrak{b} - \bar{\theta}_0^\mathfrak{c})\mathrm{e}^{-\hat{t}/\hat{\tau}_\delta},\\
 &\bar{\theta}^\mathfrak{c} = \frac{1}{\hat{c}_v^{\mathfrak{b}+\mathfrak{c}}}(\hat{c}_v^\mathfrak{b} \bar{\theta}_0^\mathfrak{b} + \hat{c}_v^\mathfrak{c} \bar{\theta}_0^\mathfrak{c})\mathrm{e}^{-\hat{t}} - \frac{\hat{c}_v^\mathfrak{b}}{\hat{c}_v^{\mathfrak{b}+\mathfrak{c}}}(\bar{\theta}_0^\mathfrak{b} - \bar{\theta}_0^\mathfrak{c})\mathrm{e}^{-\hat{t}/\hat{\tau}_\delta},
 \end{split}
 \label{relaxsol1}
\end{align}
where 
\begin{align*}
 \hat{t} = \frac{t}{\tau}, \quad \hat{\tau}_\delta = \frac{\tau_\delta}{\tau} = \frac{\tau_{\mathfrak{b}\mathfrak{c}}/ \tau}{1+\tau_{\mathfrak{b}\mathfrak{c}}/ \tau}.
\end{align*}
We have also the following relations:
\begin{align}
\begin{split}
 &\bar{\theta}^\mathfrak{b} - \bar{\theta}^{\mathfrak{b}\mathfrak{c}} = \frac{\hat{c}_v^\mathfrak{c}}{\hat{c}_v^{\mathfrak{b}+\mathfrak{c}}} (\bar{\theta}^\mathfrak{b}_0 - \bar{\theta}^\mathfrak{c}_0) \mathrm{e}^{-\hat{t}/\hat{\tau}_\delta},\\
 &\bar{\theta}^\mathfrak{c} - \bar{\theta}^{\mathfrak{b}\mathfrak{c}} = - \frac{\hat{c}_v^\mathfrak{b}}{\hat{c}_v^{\mathfrak{b}+\mathfrak{c}}} (\bar{\theta}^\mathfrak{b}_0 - \bar{\theta}^\mathfrak{c}_0) \mathrm{e}^{-\hat{t}/\hat{\tau}_\delta}.
\end{split}
 \label{relaxsol2}
\end{align}
As is expected, we can clearly see, from \eqref{relaxsol1}, \eqref{relaxsol2}, and \eqref{tbarKR}, that the temperatures $\theta^\mathfrak{b}$ and $\theta^\mathfrak{c}$ relax to the temperature $\theta^{\mathfrak{b}\mathfrak{c}}$ with the relaxation time ${\tau}_\delta$, while the  temperatures $\theta^{\mathfrak{b}\mathfrak{c}}$ and $\theta^\mathfrak{a}$ relax to the equilibrium temperature $T$ with the relaxation time ${\tau}$.

\begin{figure}[h!]
\centering
 \includegraphics[width=58mm]{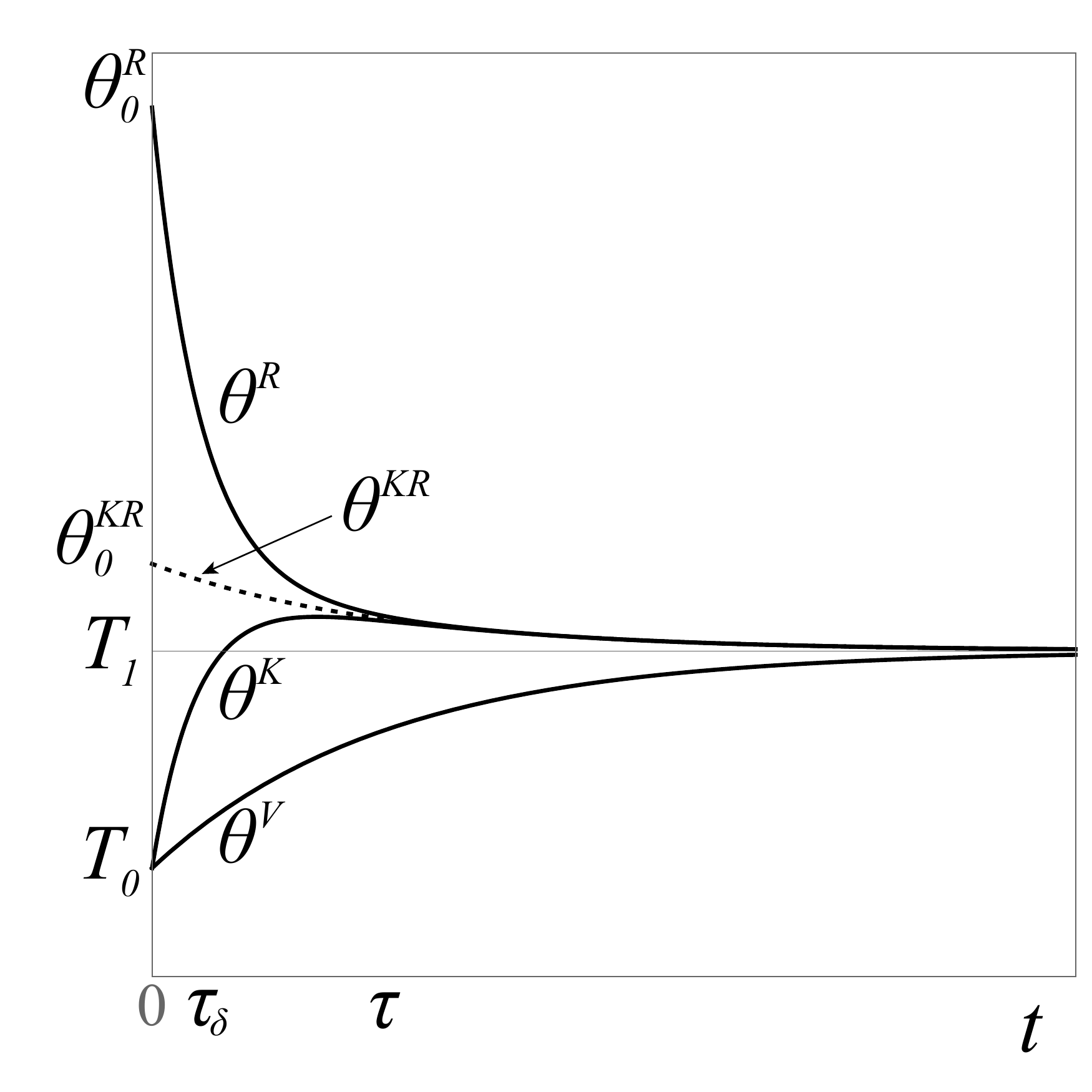} 
 \caption{Schematic time-evolution of the relaxation of the nonequilibrium temperatures $\theta^K$, $\theta^R$, $\theta^V$, and $\theta^{KR}$ in the (KR)-process. The R-mode is excited from $T_0$ to $\theta^R_0$ instantaneously at the initial time, while K and V-modes are  initially at the temperature $T_0$. Final equilibrium temperature is $T_1$. Relaxation times $\tau_{\delta}$ and $\tau$ in this case are also indicated.}
 \label{Fig:Trelax1}
\end{figure}

From experimental data on polyatomic gases such as CO$_2$, Cl$_2$, Br$_2$ gases, the (KR)-process is a suitable process \cite{Stupochenko,Kustova} (see also the analysis in Section \ref{sec:LW}).  Therefore, as a typical example, we particularly focus on this process and study the relaxation evolved from a nonequilibrium initial state: $\theta^K |_{t=0} = \theta^V|_{t=0} = T_0, \ \theta^R|_{t=0} = \theta^R_0 \ (> T_0)$.  This initial state may be generated experimentally as follows: we firstly prepare the equilibrium state with the temperature with $T_0$, then we excite only the R-mode from the temperature $T_0$ to the temperature $\theta^R_0$ instantaneously at the initial time.  The relaxation is analyzed by solving \eqref{gasatrest} under the initial condition, in which $T$ should be replaced by $T_1$.
From the condition \eqref{TET7}, $T_1$ is given by
\begin{align*}
T_1 = T_0 + \frac{c^R_v (\theta^R_0 - T_0)}{c_v} .
\end{align*}
The time-evolution of the relaxation is shown schematically in Fig.\ref{Fig:Trelax1}, from which we understand the two-step relaxation, and the energy redistribution from the R-mode to the K and V-modes.  We also notice that, after the elapse of a period of time of $O(\tau_{\delta})$ from the initial time, the relation $\theta^K = \theta^R = \theta^{KR}$ is approximately satisfied.  Therefore the results derived from ET$^{KR}_6$ and ET$_7$ with (KR)-process are nearly the same with each other. This means that ET$_7$ can be safely replaced by the simpler theory, ET$^{KR}_6$.


\section{Dispersion and attenuation of ultrasonic wave: an application of ET$_7$}\label{sec:LW}

We derive the dispersion relation of a plane harmonic wave in Section \ref{subsec:dispersion}, and discuss its general features in Section \ref{subsec:dispersion-description}.  Theoretical prediction of the attenuation per wavelength $\alpha_\lambda$ is compared with the experimental data in the case of CO$_2$ \cite{Shields-1959}, Cl$_2$ and Br$_2$ \cite{Shields-1960} gases in Section \ref{SubSec:Comparison}.

\subsection{Dispersion relation}\label{subsec:dispersion}

Let us study a plane harmonic wave propagating along the $x$-axis expressed by
\begin{align*}
 &\boldsymbol{u} = \boldsymbol{u}_0 + \bar{\boldsymbol{u}},
\end{align*}
where $\boldsymbol{u} = (\rho , v , T , \delta, \Delta)$ is a state vector
with $v$ being the $x$-component of the velocity $v_i$, and $\boldsymbol{u}_0 = (\rho_0 , 0 , T_0 , 0 ,0)$ is a state vector at a reference equilibrium state at rest. The deviation $\bar{\boldsymbol{u}}=(\bar{\rho} , \bar{v} , \bar{T} , \bar{\delta} , \bar{\Delta})$ from $\boldsymbol{u}_0$ is expressed by
\begin{align*}
 &\bar{\boldsymbol{u}}=\boldsymbol{w}\mathrm{e}^{\mathrm{i} (\omega t- k x)},
\end{align*}
where $\boldsymbol{w}$ is the amplitude vector, $\omega$ is the angular frequency, and $k$ is the complex wave number: $k=\Re (k) + \mathrm{i} \Im(k)$ being $\Re (k)$ and $\Im(k)$ the real and imaginary parts of $k$.

From the linearized system of field equations with respect to $\bar{\boldsymbol{u}}$, we obtain the dispersion relation, derivation method of which is given in \cite{RuggeriMuracchini1992}:
\begin{align*}
 &\displaystyle \frac{1}{(z {U_{Euler}})^2} =
 \left\{
\begin{array}{l}
 \displaystyle 1 + \left(\hat{U}_E^2 - \hat{U}^{\mathfrak{b}\mathfrak{c}{}^2}_E\right)\frac{\mathrm{i}\Omega \hat{\tau}_\delta}{1+ \mathrm{i}\Omega \hat{\tau}_\delta}
  +(\hat{U}^{\mathfrak{b}\mathfrak{c}{}^2}_E - 1) \frac{\mathrm{i}\Omega }{1+\mathrm{i}\Omega}  \\ \qquad    \text{for ($\mathfrak{b}\mathfrak{c}$)-process \ (($\mathfrak{b}\mathfrak{c}$)=(KR) or (KV))},\\
  \displaystyle 1 + \left(\hat{U}_E^2 - 1\right)\frac{\mathrm{i}\Omega }{1+ \mathrm{i}\Omega} \qquad \text{for (RV)-process},
 \end{array}\right.
\end{align*}
where $z \equiv  k/\omega$, $\Omega \equiv \omega \tau$, and the dimensionless characteristic velocities: $\hat{U}_{E} \equiv {U_E} / {U_{Euler}}$ and $\hat{U}_{E}^{\mathfrak{b}\mathfrak{c}} \equiv {U_E^{\mathfrak{b}\mathfrak{c}}} / {U_{Euler}}$ given by
\begin{align*}
 \hat{U}^2_E = \frac{\hat{c}_v}{\hat{c}_v^K} \frac{1+\hat{c}_v^K}{1+\hat{c}_v}, \quad
  \hat{U}^{\mathfrak{b}\mathfrak{c}{}^2}_E = \frac{\hat{c}_v}{\hat{c}_v^{\mathfrak{b}+\mathfrak{c}}} \frac{1+\hat{c}_v^{\mathfrak{b}+\mathfrak{c}}}{1+\hat{c}_v}.  
\end{align*}
For (RV)-process, $\delta$ does not play any role in the dispersion relation as seen from the linearized equations of \eqref{relax2}.

From the dispersion relation, the phase velocity $v_{{ph}}$, the attenuation factor $\alpha$, and the attenuation per wavelength $\alpha_\lambda$ are derived by using the relations:
 \begin{align*}
  \begin{split}
   &v_{ph}= \frac{\omega}{\Re (k)} , 
   \quad \alpha = -\Im (k), 
   \quad  \alpha_\lambda = \frac{2\pi v_{ph}\alpha}{\omega} = - 2 \pi \frac{\Im (k)}{\Re (k)}.
  \end{split}
 \end{align*}
In the high-frequency limit $\Omega \to \infty$, we have
\begin{align*}
 & v_{ph,\infty} \equiv \lim_{\omega \to \infty} v_{ph} = \pm {{U}_{E}},   \\
 &\alpha_\infty \equiv \lim_{\omega \to \infty} \alpha =
 \left\{
 \begin{array}{l}
  \displaystyle \pm \frac{1}{2 U_{Euler}\tau}\frac{\hat{U}_E^2 - \hat{U}_E^{\mathfrak{bc}{}^2} + \hat{\tau}_\delta (\hat{U}_E^{\mathfrak{bc}{}^2} - 1)}{\hat{U}_E^3 \hat{\tau}_\delta}  \\ \quad  \ \ \text{for ($\mathfrak{b}\mathfrak{c}$)-process (($\mathfrak{b}\mathfrak{c}$) = (KR) or (KV)),} \\
  \displaystyle \pm \frac{1}{2 U_{Euler}\tau}\frac{\hat{U}_E^2 - 1}{\hat{U}_E^3 } \ \ \ \text{for (RV)-process.}
 \end{array}
 \right
.
\end{align*}

In a similar way, we can also derive the dispersion relations of the ET$_6$ theories explained in Section \ref{sec:subsystem}, explicit expressions of which are omitted here for simplicity.  A remarkable point is as follows: the dispersion relation of ET$_6^{RV}$ \cite{ET6-LinearWave} coincides with the dispersion relation of ET$_7$ with (RV)-process.  While the dispersion relation of ET$_6^{\mathfrak{b}\mathfrak{c}} \ ((\mathfrak{b}\mathfrak{c})=(KR) \ \text{or} \ (KV))$ is obtained from the dispersion relation of ET$_7$ with ($\mathfrak{b}\mathfrak{c}$)-process by taking the limit $\tau_\delta \to 0$.

\subsection{Qualitative description of the dispersion relation} \label{subsec:dispersion-description}

In this subsection, we discuss the general features of the dispersion relation by studying some typical cases
so that we may address the following two questions: (i)  For given experimental data, how can we determine the most suitable relaxation process among possible (KR), (KV), (RV)-processes?  (ii) What is the relationship between the applicability ranges of ET$_7$ and ET$_6$ theories?

\begin{figure}[h!]
	\centering
	\includegraphics[width=80mm]{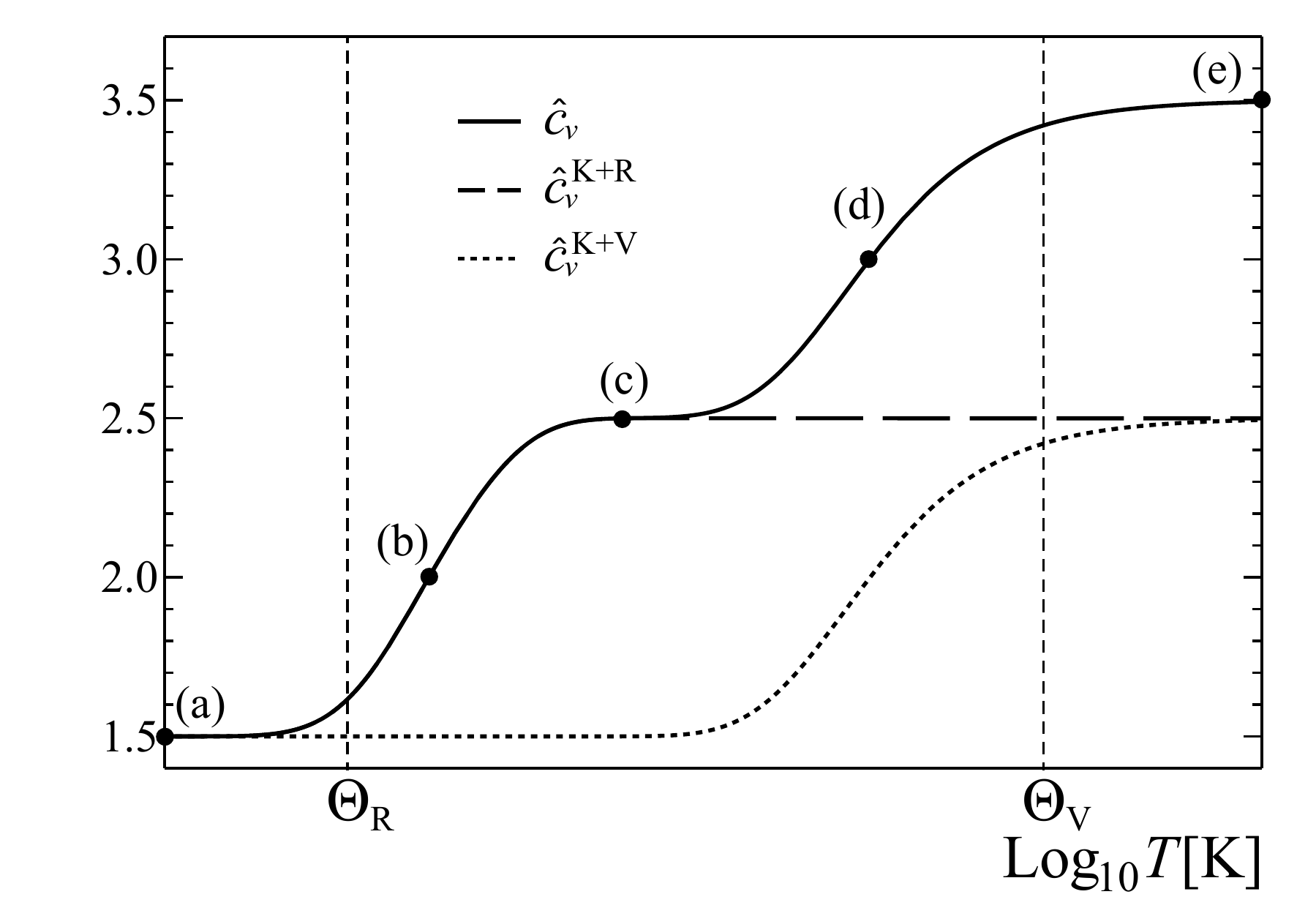} 
	\caption{Typical temperature dependence of the dimensionless specific heats; $\hat{c}_v$, $\hat{c}_v^{K+R}$, and $\hat{c}_v^{K+V}$. The five cases (a)-(e) listed in Table \ref{FiveCases} are also indicted.}
	\label{Fig:cvH22}
\end{figure}

In the above, we have noticed that the dispersion relation depends on the temperature through the specific heats. Therefore, before going into the main discussions, we remark here on the estimation method of the specific heats. As usual in thermodynamics, we may use the experimental data on the specific heats.  However, for a simple gas like a homonuclear diatomic molecule gas, which we adopt in this subsection, the specific heats can be estimated by the statistical-mechanical considerations. 
That is, the specific heats $c_v^R$ and $c_v^V$ are evaluated by using the rotational and vibrational partition functions $Z^R$ and $Z^V$ as follows:
\begin{align}
 \begin{split}
 &c_v^R(T) = \frac{k_B}{m} \beta_E^2 \frac{\partial^2 Z^{R}}{\partial \beta_E^2},
  \quad \text{with} \quad
  Z^R = Z_g^{\frac{s_n}{2s_n+1}} Z_u^{\frac{s_n+1}{2s_n+1}},\\
  &Z_g=\sum_{l=\text{even}}^\infty (2l+1) \mathrm{e}^{-k_B \Theta_R \beta_E l (l+1)},\ \ 
  Z_u=\sum_{l=\text{odd}}^\infty (2l+1) \mathrm{e}^{-k_B \Theta_R \beta_E l (l+1)},\\
  &c_v^V(T) = \frac{k_B}{m} \beta_E^2 \frac{\partial^2 Z^{V}}{\partial \beta_E^2},
 \quad \text{with} \quad
Z^{V} = \prod_{i=1}^{N}\frac{\mathrm{e}^{- k_B \Theta_{V_i} \beta_E/2}}{1 - \mathrm{e}^{- k_B \Theta_{V_i} \beta_E}},  
 \end{split}
 \label{cvSM}
\end{align}
where $s_n$, $\Theta_R$, and $\Theta_{V_i}$ are, respectively, the nuclear spin, the characteristic rotational temperature, and vibrational temperature of the $i$-th harmonic mode. In the case of diatomic molecules with $s_n=1/2$ and $N=1$, a typical temperature dependence of the specific heats $c_v$, $c_v^{K+R}$, and $c_v^{K+V}$ is shown in Fig. \ref{Fig:cvH22}.  

Let us study the temperature dependence of the phase velocity $v_{ph}(\omega)$ and the attenuation per wavelength $\alpha_\lambda(\omega)$ in the five typical cases listed in Table \ref{FiveCases}.  The temperature of the reference equilibrium state $\boldsymbol{u}_0$ increases from the case (a) to the case (e) as seen in Fig. \ref{Fig:cvH22}.  

\begin{table}[h!]
	\centering
	\caption{Five typical cases.  Translational mode is fully excited in all cases.}
 {
 \begin{ruledtabular}
		\begin{tabular}{clll}
			Case &  Specific heats  & Rotational mode & Vibrational mode  \\ \hline 
			 (a) & $\hat{c}_v^K = 3/2$, $\hat{c}_v^{R} = 0$, $\hat{c}_v^{V} = 0$  & Ground state & Ground state  \\ 
			 (b) & $\hat{c}_v^K = 3/2$, $\hat{c}_v^{R} = 1/2$, $\hat{c}_v^{V} = 0$  & Partly excited & Ground state \\  
			 (c) &  $\hat{c}_v^K = 3/2$, $\hat{c}_v^{R} = 1$, $\hat{c}_v^{V} = 0$  & Fully excited & Ground state \\  
			 (d) &  $\hat{c}_v^K = 3/2$, $\hat{c}_v^{R} = 1$, $\hat{c}_v^{V} = 1/2$  & Fully excited & Partly excited \\  
			 (e) & $\hat{c}_v^K = 3/2$, $\hat{c}_v^{R} = 1$, $\hat{c}_v^{V} = 1$  & Fully excited & Fully excited  
		\end{tabular}
  \end{ruledtabular}
	}\label{FiveCases}
\end{table}

As many experimental data \cite{Bhatia} indicate that the ratio of the relaxation times $\hat{\tau}_\delta$ is $O(10^{-3})$ or more (see Section \ref{SubSec:Comparison} for (KR)-process), we assume here that $\hat{\tau}_\delta = 10^{-3}$ for all processes. Therefore we can observe the slow and fast relaxation processes separately. In fact, we expect that the dispersion relation has a remarkable change at around $\Omega \sim O(1) \ (\omega \sim O(\tau^{-1}))$ and $\Omega \sim O(10^3) \ (\omega \sim O(\tau_\delta^{-1}))$. See also \textit{Remark \ref{R:LWtau}} below.

In Fig.\ref{Fig:v}, the dimensionless phase velocity $\hat{v}_{ph}=v_{ph}/{U_{Euler}}$ and the attenuation per wavelength $\alpha_\lambda$ predicted by ET$_7$ with (KR), (KV) and (RV)-processes in the five cases (a)-(e) are shown.
\begin{figure}[h!]
	\centering
	\includegraphics[width=86mm]{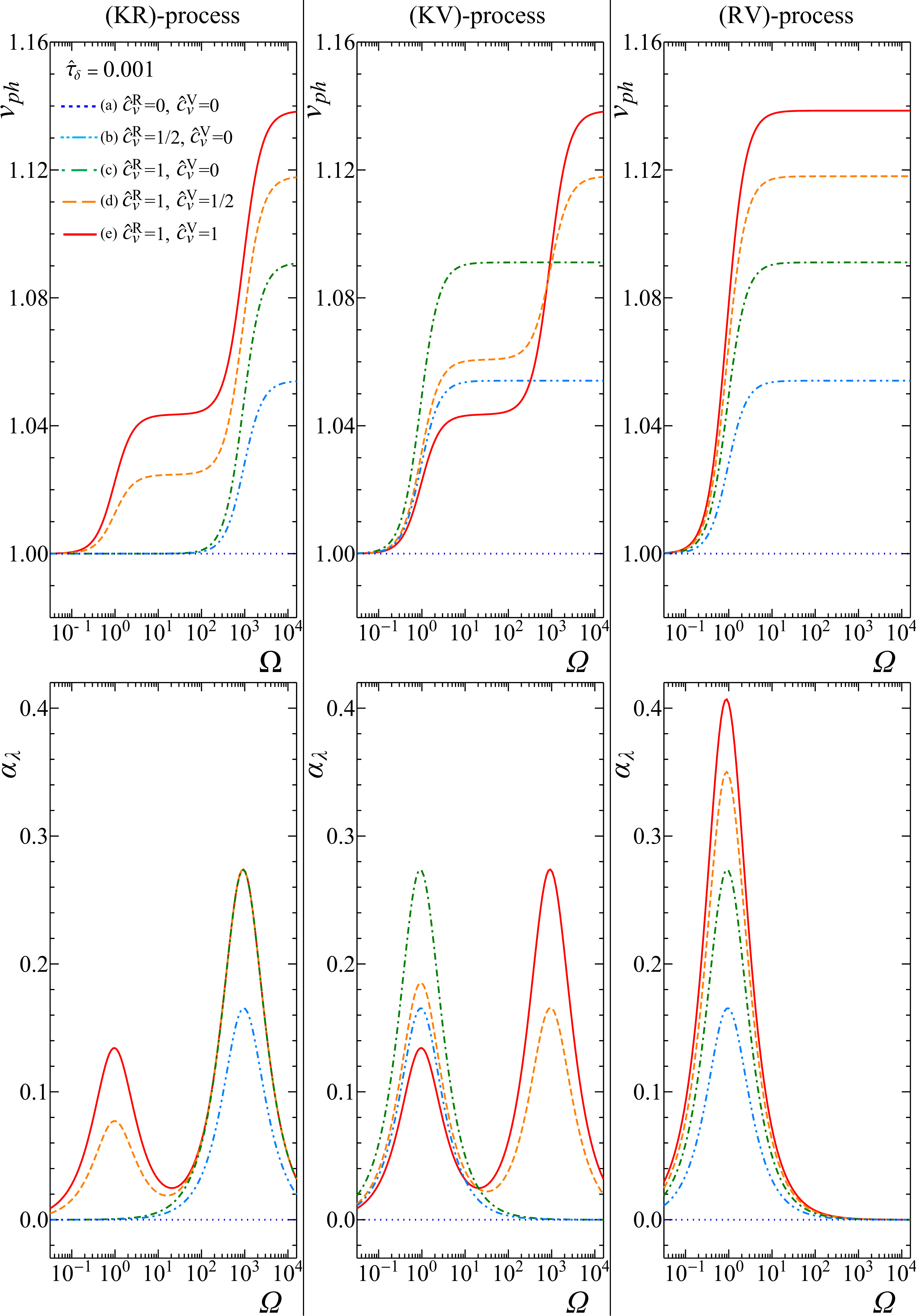}
	\vspace{-2mm}
	\caption{Dimensionless phase velocity $\hat{v}_{ph}=v_{ph}/{U_{Euler}}$ and the attenuation per wavelength $\alpha_\lambda$ for the (KR), (KV), and (RV)-processes in the five cases (a)-(e) listed in Table \ref{FiveCases}. The ratio of the relaxation time $\hat{\tau}_\delta$ is $10^{-3}$.}
	\label{Fig:v}
\end{figure}
Noticeable points are summarized as follows:
\begin{enumerate}
\item Among the three relaxation processes, i.e., (KR), (KV), and (RV)-processes, the dependence of the curve $v_{ph}(\omega)$ on the temperature is quite different from each other. In other words, each relaxation process has its own characteristic temperature dependence of the curve $v_{ph}(\omega)$.  Conversely, experimental data on such a temperature dependence can afford a suitable method to identify the relaxation process in a gas under study.
\item The dependence of the curve $\alpha_\lambda(\omega)$ on the temperature is also quite different from each other among the three relaxation processes. Experimental data on such a temperature dependence can afford another suitable method to identify the relaxation process in a gas under study.  To be more precise, let us focus on the temperature dependence of the value of $\alpha_\lambda$ at its peak in the low frequency region, i.e., the left peak in Fig. \ref{Fig:v}.  The peak value $\alpha_\lambda^{peak}$ attained at $\Omega = \Omega^{peak}$ is explicitly given by
\begin{align*}
	&\alpha_\lambda^{peak} = 2\pi \frac{{\hat{U}_{E}^{\mathfrak{b}\mathfrak{c}}}-1}{{\hat{U}_{E}^{\mathfrak{b}\mathfrak{c}}}+1} \ \ \ \ \  \left(\Omega^{peak} = \frac{1}{{\hat{U}_{E}^{\mathfrak{b}\mathfrak{c}}}}\right) \\ &((\mathfrak{b}\mathfrak{c})=(KR), (KV), (RV)),
\end{align*}
and its temperature dependence is shown in Fig. \ref{Fig:cvtemp}.
	 \begin{figure}[h!]
	 	\centering
	 	\includegraphics[width=80mm]{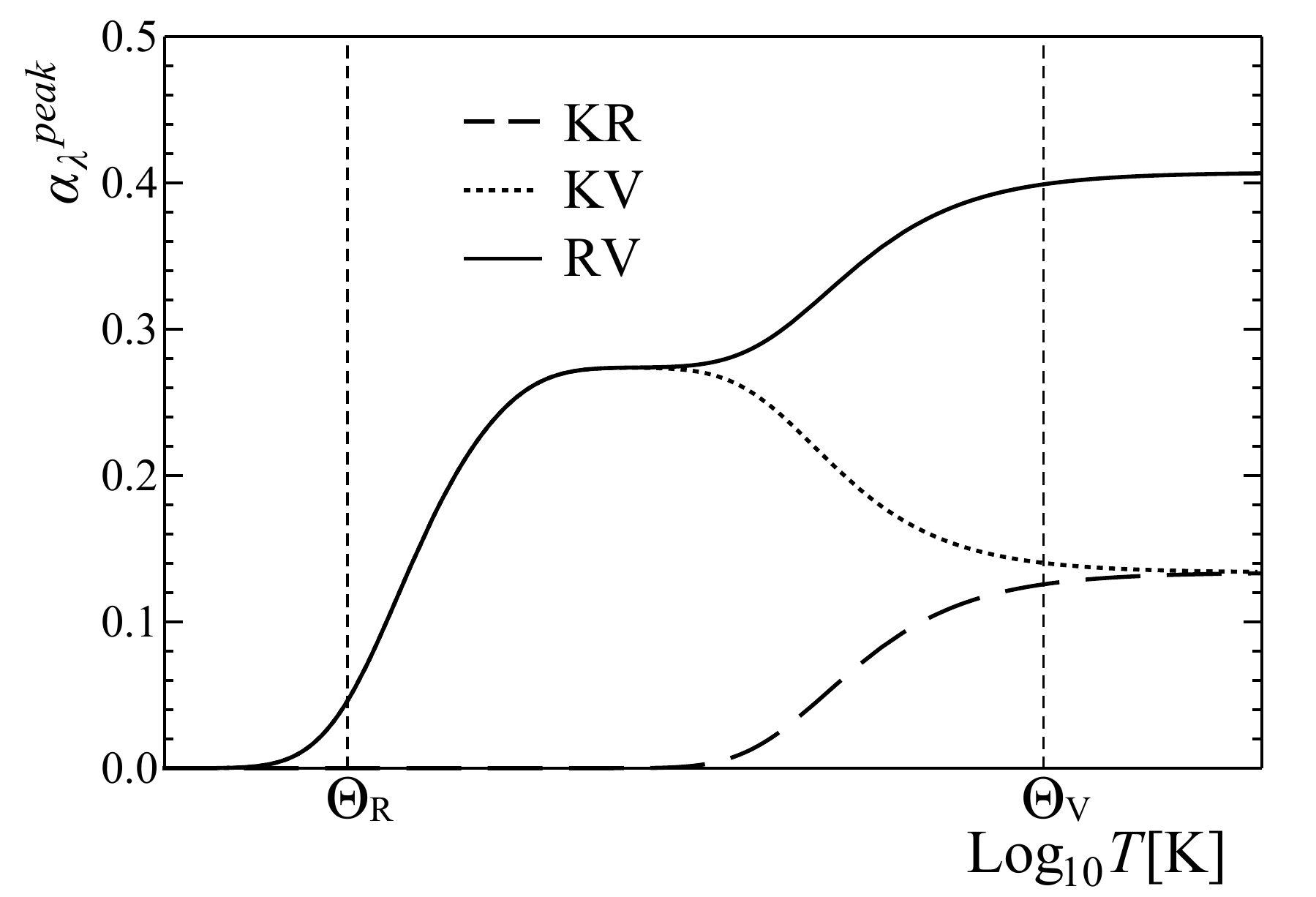}
	 	\vspace{-2mm}
	 	\caption{Temperature dependence of $\alpha_\lambda^{peak}$ for (KR), (KV) and (RV)-processes. The characteristic rotational and vibrational temperatures are denoted as $\Theta_R$ and $\Theta_V$.}
	 	\label{Fig:cvtemp}
	 \end{figure}
\item From the remark about the dispersion relation of ET$_6$ in Section \ref{subsec:dispersion} and the curves of ET$_7$ shown in Fig.\ref{Fig:v}, we conclude that the ET$_6$ theories are reliable in the frequency region $\Omega < O(10^1)$, where ET$_6$ theories are quite good approximation of the ET$_7$ theory.  It should be emphasized that even in this frequency region we should pick up a suitable ET$_6$ theory among the three theories by using the methods mentioned above.  When we go into higher frequency region $\Omega > O(10^1)$, we should adopt the the ET$_7$ theory instead of the ET$_6$ theory.  This is true especially for (KR) and (KV)-processes.  	
	
\end{enumerate}

To sum up, we have proposed the selection methods for the most suitable relaxation process and made clear the applicability ranges of ET$_6$ and ET$_7$ theories.  Finally we point out that the general features of the dispersion relation discussed above can be found not only diatomic gases but also in polyatomic gases because such features come mainly from the global dependence of the specific heats on the temperature.

\medskip

\begin{figure}[h!]
	\begin{center}
		\includegraphics[width=80mm]{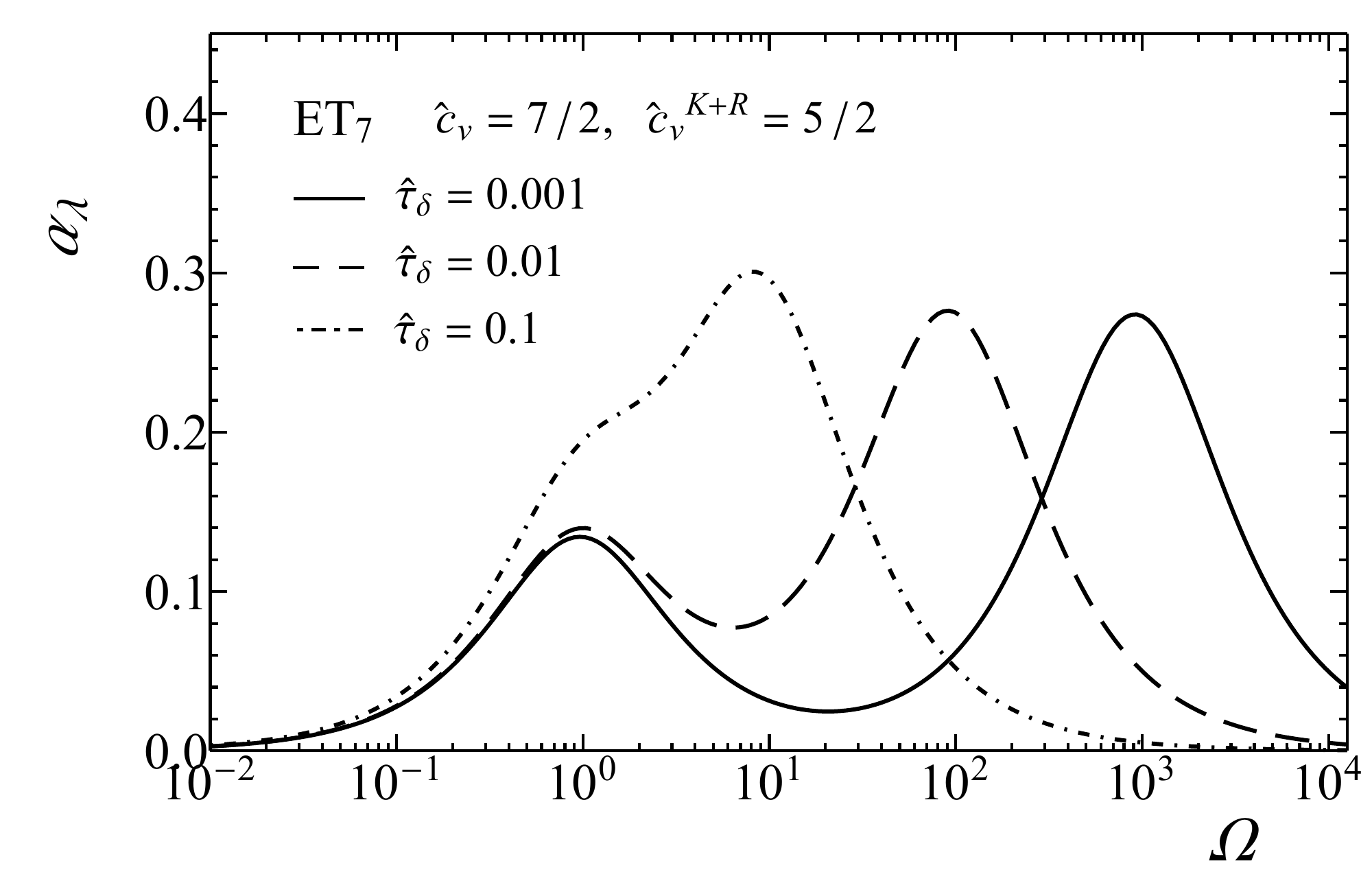} 
	\end{center}
	\vspace{-7mm}
	\caption{Dependence of $\alpha_\lambda$ for (KR)-process on $\Omega$ with $\hat{c}_v = 7/2$ and $\hat{c}_v^{K+R} = 5/2$. The solid, dashed and dotted lines indicate, respectively, the cases with $\hat{\tau}_\delta = 0.001, 0.01$ and $0.1$.}
	\label{Fig:al-t}
\end{figure}

\Remark[\label{R:LWtau}]
In Fig.\ref{Fig:al-t}, we show the dependence of $\alpha_\lambda$ on the ratio of the relaxation times $\hat{\tau}_{\delta}$ in the case of the (KR)-process for an example.  We notice from the figure that, when the ratio increases, two peaks gradually coalesce into a big one.  In such a case the prediction of ET$_6^{KR}$ is no longer valid even in the frequency region $\Omega < O(10^1)$, and ET$_7$ should be used.

\medskip

\Remark[\label{LWclassical}]
As explained in Section \ref{sec:ET7}, the ET$_7$ theory neglects the so-called classical absorption, that is, the attenuation due to the shear viscosity and the heat conduction. For gases in which these effect emerges in the higher frequency region $\Omega >> O(10^1)$ such as H$_2$ and CO$_2$ \cite{ET14linear,ET14shock},  there is a possibility 
that two peaks from this and from the rapid relaxation studied above, i.e., the right peak in Fig. \ref{Fig:v} coalesce into one. Because of this, in the above, we have focused on the temperature dependence of the peak value of $\alpha_\lambda$ in the low frequency region.
On the other hand, there is another possibility: for gases in which the effects of shear viscosity and heat conduction emerge around $\Omega \sim O(1)$, the peaks from this and from the left peak in Fig. \ref{Fig:v} coalesce into one. Moreover, if $\hat{\tau}_\delta \sim O(1)$, all three peaks coalesce into one.
In the next paper, we will study such a combined effects in detail.

\subsection{Comparison with experimental data} \label{SubSec:Comparison}

We compare the theoretical prediction of $\alpha_\lambda$ by ET$_7$ with the experimental data of CO$_2$ \cite{Shields-1959}, Cl$_2$ and Br$_2$ gases \cite{Shields-1960}.

\begin{figure}[h!]
\begin{center}
 \includegraphics[width=80mm]{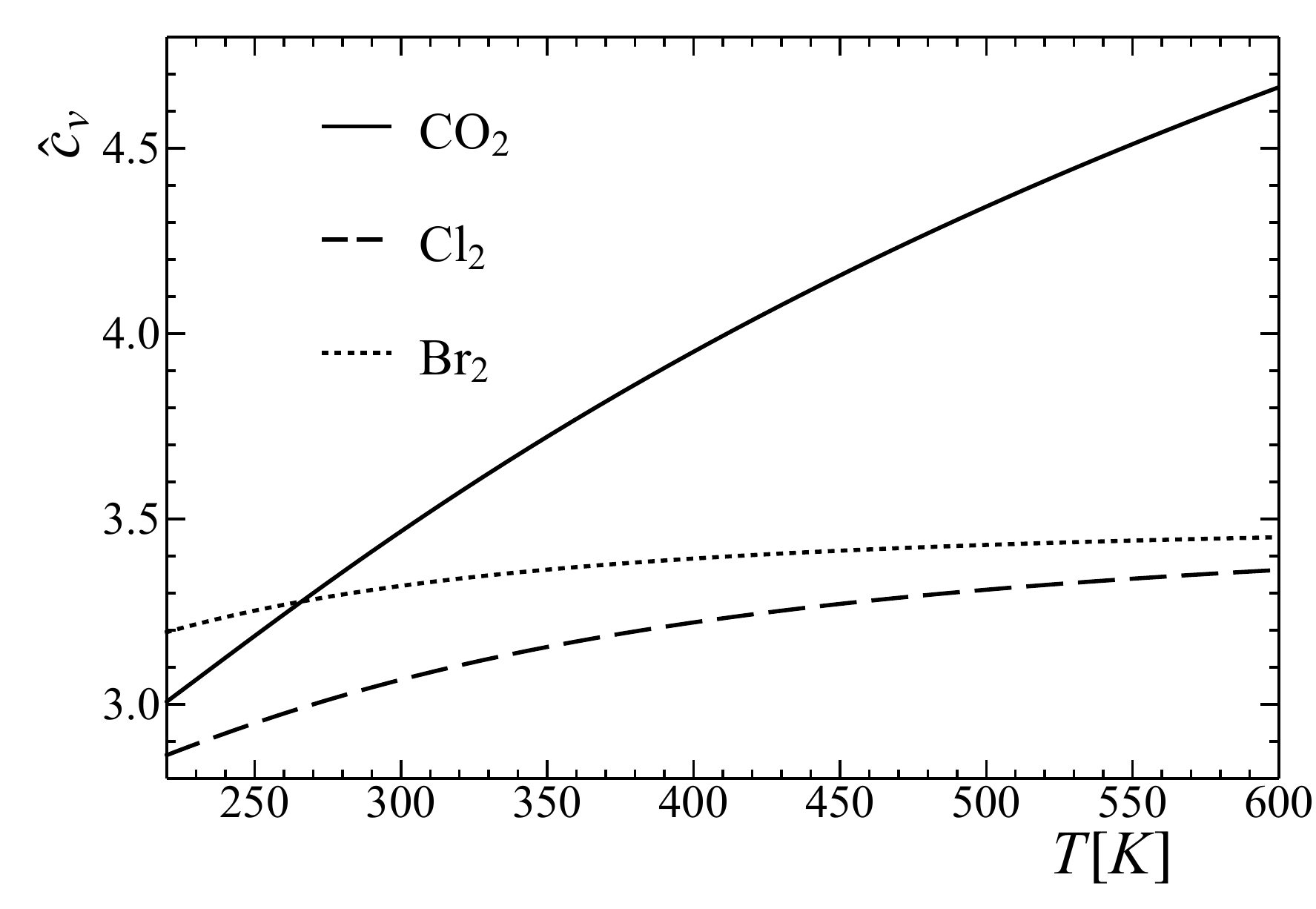} 
\end{center}
  \vspace{-9mm}
 \caption{Dependence of $\hat{c}_v$ on $T$.}
 \label{Fig:cv}
\end{figure}

As a preliminary step, we evaluate the specific heats of CO$_2$, Cl$_2$, and Br$_2$ gases by the  statistical-mechanical method.
In these gases, the characteristic rotational temperature $\Theta_R$ is very low. In fact, from the data on the rotational constant at  the ground state \cite{Radzig}, it is estimated as $0.56$K for CO$_2$, $0.35$K for Cl$_2$, and $0.12$K for Br$_2$. Therefore, in the temperature range higher than the room temperature, the rotational degrees of freedom of these gases are in a fully excited state with $\hat{c}_v^{K+R} = 5/2$.
While the temperature dependence of the vibrational specific heat is approximately calculated by \eqref{cvSM}.
For CO$_2$ molecule with $N=4$, the characteristic vibrational temperatures are given by $\Theta_{V_1} = \Theta_{V_2} = 960$K, $\Theta_{V_3} = 1997$K and $\Theta_{V_4} = 3380$K \cite{Radzig}.
For Cl$_2$ and Br$_2$ molecules with $N=1$, the characteristic vibrational temperatures are, respectively, $\Theta_{V} = 805$K and $\Theta_{V} =468$K \cite{Radzig}.  The temperature dependence of $\hat{c}_v$ is shown in Fig.\ref{Fig:cv}.

Applying the selection method mentioned above to the experimental data on $\alpha_\lambda$ \cite{Shields-1959,Shields-1960}, we conclude that these gases have the (KR)-process and the relaxation time $\tau$ is several orders larger than the relaxation time $\tau_\delta$.  Therefore, as the present comparison is made only in the low frequency region, we may safely assume $\hat{\tau_\delta} = 10^{-3}$. 

As the experimental data are summarized as the relationship between $\alpha_\lambda$ and $f/p$ [Hz/Pa] ($f=\omega/2\pi$) \cite{Shields-1959,Shields-1960}, we use the quantity $\omega/p$ instead of $\Omega$. Recalling that $\Omega =  (\tau p) ({\omega}/{p})$, we adopt the quantity $\tau p$ as a fitting parameter determined by the least square method. 

\begin{figure}[h]
 \includegraphics[width=74mm]{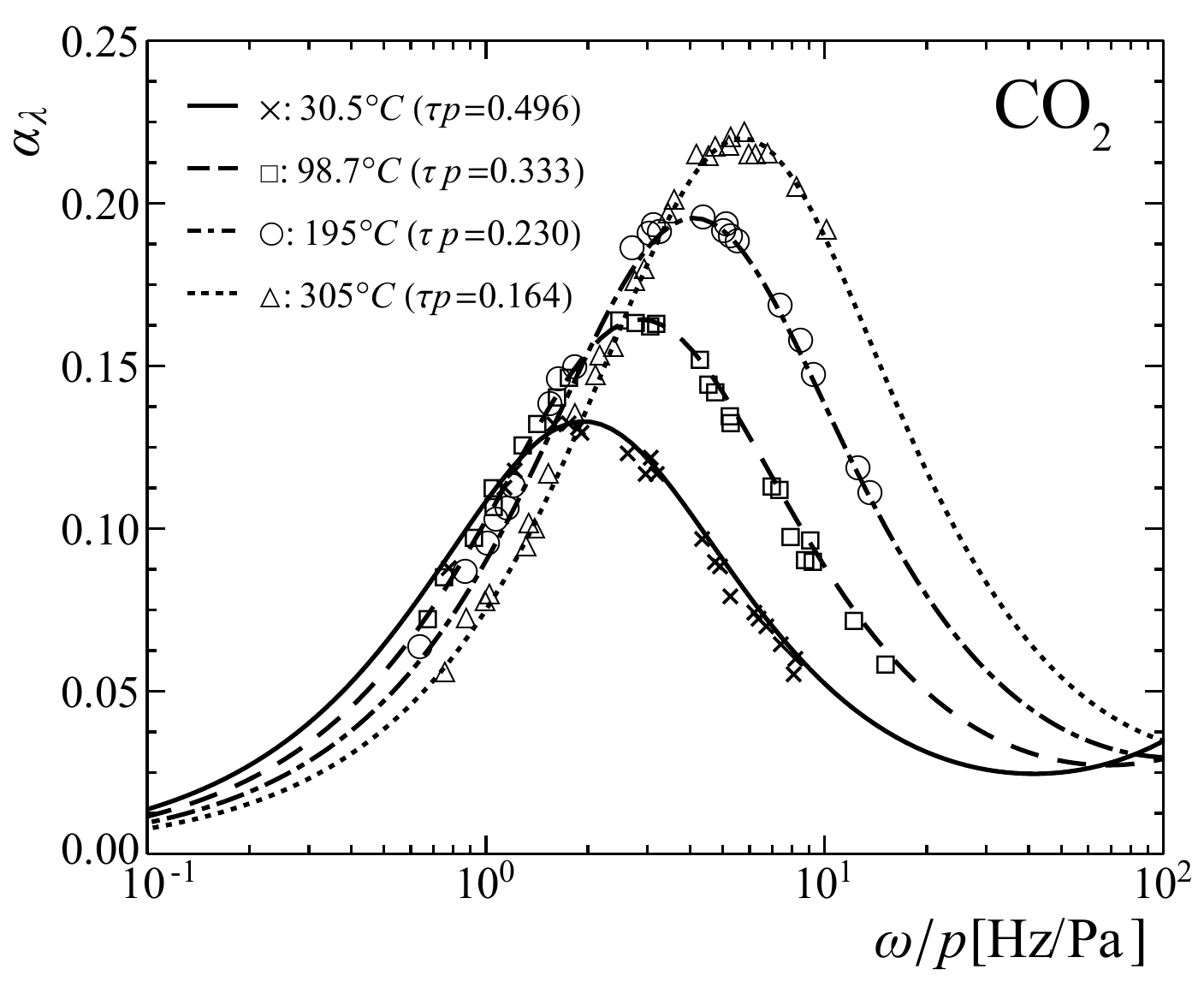} 
 \includegraphics[width=74mm]{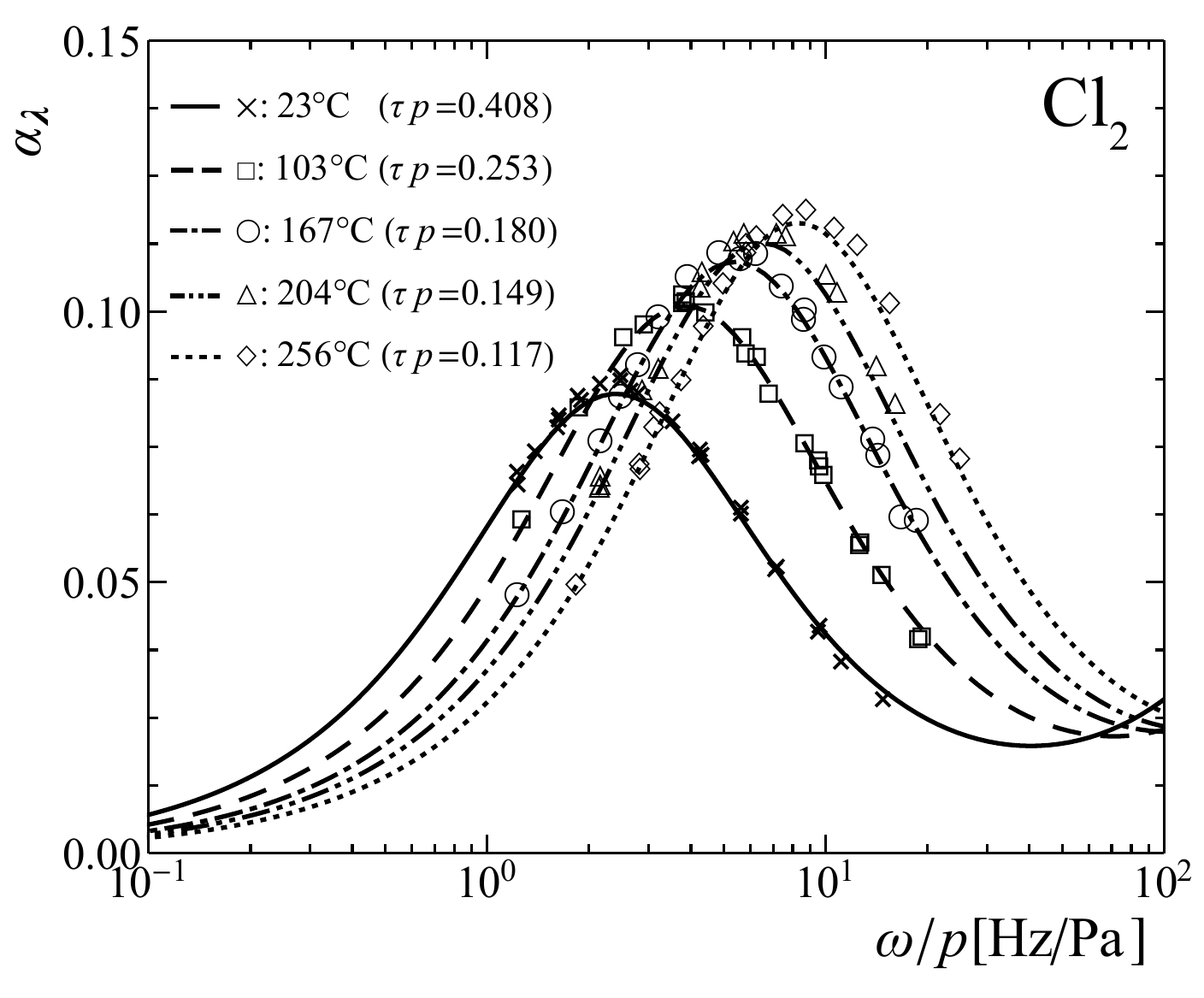}
 \includegraphics[width=74mm]{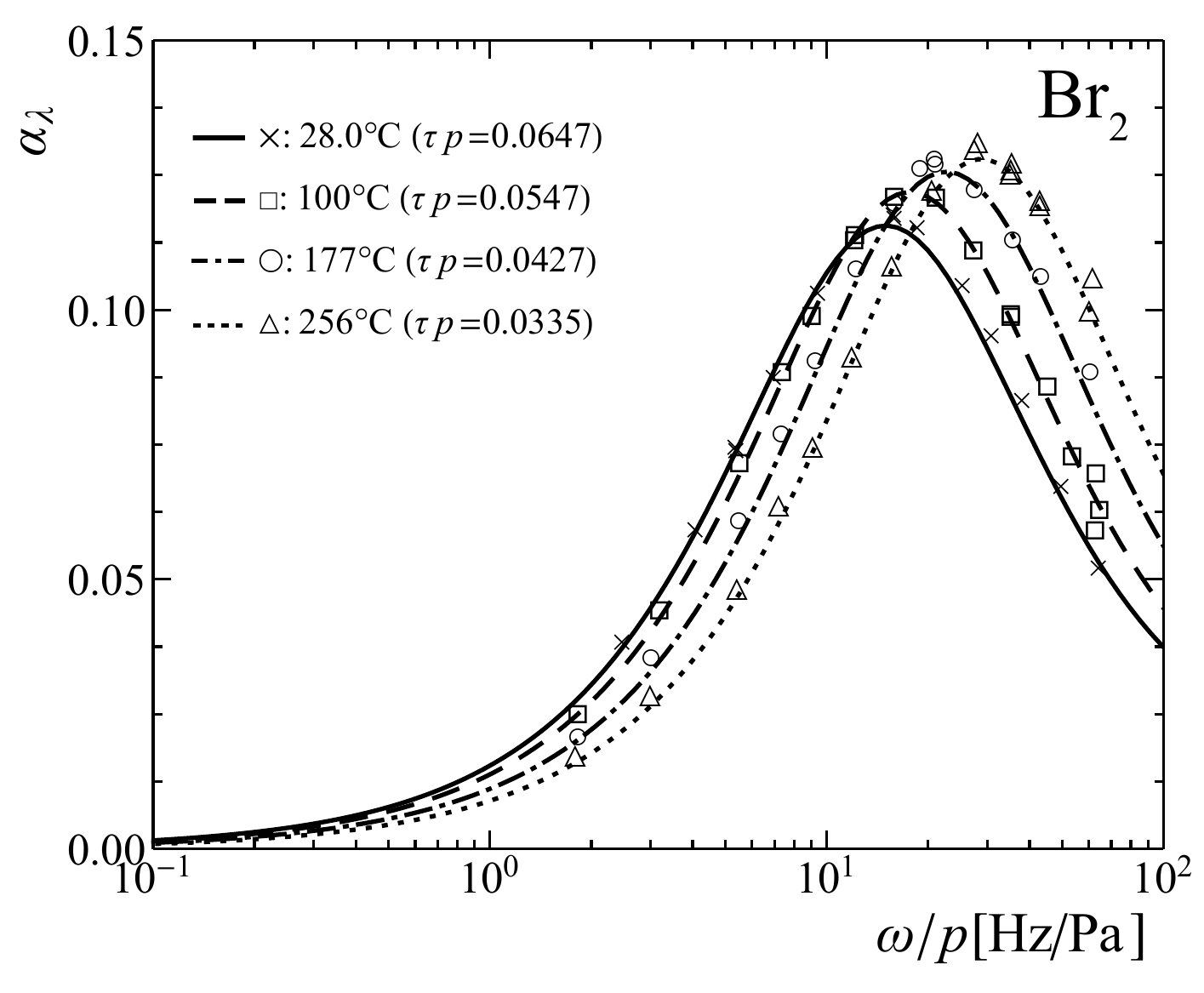}
 \caption{Dependence of $\alpha_\lambda$ on $\omega/p$ [Hz/Pa] for several temperatures with $\hat{c}_v^{K+R} = 5/2$, $\hat{\tau_\delta} = 10^{-3}$ in rarefied CO$_2$, Cl$_2$, and Br$_2$ gases \cite{Shields-1959,Shields-1960}. A parameter $p \tau$ is chosen to fit the experimental data by the least square method.}
 \label{Fig:CO2}
\end{figure}

The comparison is made in Fig.\ref{Fig:CO2}. These figures show the excellent agreement between the theoretical prediction of ET$_7$ and the experimental data. The selected parameter $\tau p $ and the bulk viscosity $\nu^{V}$ estimated by using \eqref{nuv} are summarized in Table.\ref{ptd}.  We also emphasize the importance of the dynamic pressure in the wave propagation phenomena. This is because the bulk viscosity coefficients of CO$_2$, Cl$_2$, and Br$_2$ gases are much larger than the shear viscosity coefficients that are estimated as $1.49\times 10^{-5}$[Pa$\cdot$s] for CO$_2$, $1.363\times 10^{-5}$[Pa$\cdot$s] for Cl$_2$, and $9.42\times 10^{-4}$[Pa$\cdot$s] for Br$_2$ at $T=298$K and $p=1$atm \cite{JSME}.

\begin{table}[h]
 \centering
 \caption{The parameter $\tau p$ and the bulk viscosity.}
 \begin{ruledtabular}
 \begin{tabular}{cccc} 
  Gas& $T$ [${}^\circ\mathrm{C}$] & $\tau p$ [Pa $\cdot$ s] &$\nu^{V}$ [Pa $\cdot$ s]\\ \hline 
  CO$_2$ & 30.5& $4.96 \times 10^{-1}$ & $5.61 \times 10^{-2}$\\
  & 98.7 & $3.33 \times 10^{-1}$ & $4.62 \times 10^{-2}$\\
  & 195 & $2.30 \times 10^{-1}$ & $3.75 \times 10^{-2}$\\
  & 305 & $1.64 \times 10^{-1}$ & $2.99  \times 10^{-2}$\\ \hline
  Cl$_2$  & 23 & $4.08 \times 10^{-1}$ & $ 2.98\times 10^{-2}$\\
    & 103 & $2.53 \times 10^{-1}$ & $2.19 \times 10^{-2}$\\
    & 167 & $1.80 \times 10^{-1}$ & $1.68 \times 10^{-2}$\\
    & 204 & $1.49 \times 10^{-1}$ & $1.43 \times 10^{-2}$\\
    & 256 & $1.17 \times 10^{-1}$ & $1.16 \times 10^{-2}$\\ \hline
  Br$_2$  & 28.0 & $6.47 \times 10^{-2}$& $6.40 \times 10^{-3}$ \\
    & 100 & $5.47 \times 10^{-2}$ & $5.69 \times 10^{-3}$\\
    & 177 & $4.27 \times 10^{-2}$ & $4.57 \times 10^{-3}$\\
    & 256 & $3.35 \times 10^{-2}$ & $3.65 \times 10^{-3}$\\ 
 \end{tabular}    
 \end{ruledtabular}
 \label{ptd}
\end{table}

\medskip

\Remark
Many studies of the dispersion relation of sound in polyatomic gases have been made basing on nonequilibrium thermodynamics and/or the kinetic theory \cite{HerzfeldRice,Kneser,Bourgin1}. Except for different definitions of the relaxation times, these theories equally describe well the absorption of sound due to the energy exchange among the degrees of freedom of a molecule up to some limited frequency  \cite{Mason,Bhatia} (see also \cite{Markham} for the classification of the previous studies). In particular, the Meixner theory with the relaxation processes of the molecular internal energies \cite{Meixner-1943,Meixner-1952} has been used to describe the attenuation of sound phenomenologically.  As shown in the present paper, by using the correspondence relationship between the Meixner theory and the ET$_7$ theory discussed in Section \ref{subsec:Meixner}, the Meixner theory seems to be valid also for phenomena out of local equilibrium to which ET$_7$ is applicable.  However, as remarked above, in the high frequency region where shear viscosity and heat conduction play roles, the ET theory with more independent variables becomes to be indispensable because there exists no such correspondence relationship.

\section{Summary and outlook}\label{sec:summary}

The ET theory of rarefied polyatomic gases with two molecular relaxation processes for the rotational and vibrational modes has been constructed.  We have introduced the generalized BGK model for the collision term. After discussing the general structure of the ET theory with the triple hierarchy, we have established, in particular, the ET$_7$ theory. This theory includes three six-field theories as special cases depending on the molecular collisional process.  Finally, as an application of the ET$_7$ theory, the dispersion relation of ultrasonic wave has been derived, and excellent agreement between its theoretical prediction and the experimental data of CO$_2$, Cl$_2$, and Br$_2$ gases has been confirmed.

In our plan, the present paper is the first one in a series of papers.  We will report the following studies:
(i) As mentioned above, by using the triple hierarchy, more sophisticated ET theory including also the shear stress and heat flux as independent variables will soon be reported.  
(ii) In linear waves, the excitations of the translational, rotational, and vibrational modes from a reference state are small.  However, the ET theory can be also applied to the phenomena in which large excitations take place.  In this respect, shock wave phenomena is worth studying.  In \cite{ET6shock}, peculiar shock wave structure in a polyatomic gas was studied by the ET$_6$ theory.  When we analyze the shock wave phenomena by the present ET$_7$ theory, we can find a more detailed shock wave structure, in particular, in the relaxation region after the subshock.  
(iii) The ET theory of dense polyatomic gases with two molecular relaxation processes will also be constructed by using the duality principle developed in \cite{denseET}.

\section*{Acknowledgments}
This work was partially supported by JSPS KAKENHI Grant Numbers JP15K21452 (T.A.)  and by National Group of Mathematical Physics GNFM-INdAM (T.R.).

\appendix*
\section{Proof of Statement 2}
Let us introduce the Lagrange multipliers $\{\lambda, \ \lambda_i, \ \mu^K(\equiv\lambda_{ll}/3), \ \mu^R, \ \mu^V\}$ that correspond to the densities $\{F, F_i, F_{ll}, H_{ll}^R, H_{ll}^V\}$. The velocity dependence of the Lagrange multipliers \eqref{lam-v} is explicitly expressed as follows:
\begin{align*}
 &\lambda = \hat{\lambda} - \hat{\lambda}_i v_i +\hat{\mu}^K v^2, \quad \lambda_i = \hat{\lambda}_i - 2 \hat{\mu}^K v_i, \\
 &\mu^K = \hat{\mu}^K, \quad \mu^R = \hat{\mu}^R , \quad \mu^V = \hat{\mu}^V.
\end{align*}

From \eqref{fNML}, it is possible to express the distribution function of the truncated system \eqref{ET7} as follows
\begin{align}
 f^{(7)} = \Omega \mathrm{e}^{-\eta_i C_i}\mathrm{e}^{- \beta^K\frac{m C^2}{2}}\mathrm{e}^{-\beta^R I^R}\mathrm{e}^{-\beta^V I^V}, \label{f70}
\end{align}
where
\begin{align*}
 &\Omega = \exp \left(-1-\frac{m}{k_B}\hat{\lambda}\right), \ \ \eta_i = \frac{m}{k_B}\hat{\lambda}_i,  \\
 &\beta^K = \frac{2}{k_B} \hat{\mu}^K, \ \ \beta^R = \frac{2}{k_B} \hat{\mu}^R, \ \ \beta^V = \frac{2}{k_B} \hat{\mu}^V. 
\end{align*}
In addition, we introduce the following three parameters $\theta^K$, $\theta^R$ and $\theta^V$ through $\beta^K$, $\beta^R$, and $\beta^V$ as follows:
  \begin{align*}
   \theta^K = \frac{1}{k_B \beta^K} , \quad \theta^R = \frac{1}{k_B \beta^R}, \quad \theta^V = \frac{1}{k_B \beta^V}.
  \end{align*}
Recalling \eqref{AEi} and substituting \eqref{f70} into \eqref{field7} evaluated at zero velocity, we obtain $\eta_i=0$ and
 \begin{align*}
  \begin{split}
	&\rho = m \left(\frac{2\pi k_B \theta^K}{m}\right)^{3/2} A^R(\theta^R) A^V(\theta^V)  \Omega,\\
   &\varepsilon^K(\theta^K) = \frac{3}{2}\frac{k_B}{m}\theta^K, \\
  &\varepsilon^R(\theta^R) = \frac{k_B}{m}{\theta^R}^2\frac{\mathrm{d} \log A^R(\theta^R)}{\mathrm{d}\theta^R} ,\\
  & \varepsilon^V(\theta^V) = \frac{k_B}{m}{\theta^V}^2 \frac{\mathrm{d} \log A^V(\theta^V)}{\mathrm{d}\theta^V}.
  \end{split}
 \end{align*}
These indicate that $\theta^K$, $\theta^R$, and $\theta^V$ are the nonequilibrium temperatures of K, R and V-modes, respectively.
Then $\Omega$, $\beta^K$, $\beta^R$ and $\beta^V$ are expressed in terms of $\rho$, $\theta^K$, $\theta^R$ and $\theta^V$ as follows:
  \begin{align*}
	& \Omega = \frac{\rho}{m A^R(\theta^R) A^V(\theta^V)}\left(\frac{m}{2\pi k_B \theta^K}\right)^{3/2},\\
   &\beta^K=\frac{1}{k_B \theta^K}, \quad \beta^R=\frac{1}{k_B \theta^R} , \quad \beta^V = \frac{1}{k_B \theta^V},
  \end{align*}
and we finally obtain the nonequilibrium distribution function \eqref{eq:f7}.

The Lagrange multipliers are expressed in terms of the independent fields as follows:
  \begin{align*}
   \hat{\lambda} = -\frac{k_B}{m}\left(1+\log \Omega\right), \ \ \hat{\mu}^K = \frac{1}{2\theta^K}, \ \ \hat{\mu}^R = \frac{1}{2\theta^R}, \ \ \hat{\mu}^V = \frac{1}{2\theta^V}.
  \end{align*}
Recalling \eqref{sE-rare} with \eqref{def-g}, we obtain the following relations:
 \begin{align*}
  &\frac{k_B}{m}\log\left[\frac{m}{\rho}\left(\frac{2\pi k_B \theta^K}{m}\right)^{3/2}\right] = s^K_E(\rho,\theta^K)  - \frac{\varepsilon^K_E(\theta^K)}{\theta^K} = - \frac{g^K_E(\rho,\theta^K)}{\theta^K}, \\
 &\frac{k_B}{m}\log A^R(\theta^R) = s^R_E(\theta^R)  - \frac{\varepsilon^R_E(\theta^R)}{\theta^R} = - \frac{g^R_E(\theta^R)}{\theta^R}, \\
  &\frac{k_B}{m}\log A^V(\theta^V) = s^R_E(\theta^V)  - \frac{\varepsilon^V_E(\theta^V)}{\theta^V} = - \frac{g^V_E(\theta^V)}{\theta^V}, 
 \end{align*}
and the relations \eqref{mainfield} have been derived.



\end{document}